\documentclass[10pt,twoside,british,english]{article}
\usepackage{lmodern}

\usepackage[T1]{fontenc}
\usepackage[utf8]{inputenc}
\usepackage[a4paper]{geometry}
\usepackage{color}
\usepackage{babel}
\usepackage{float}
\usepackage{amsthm}
\usepackage{amsmath}
\usepackage{amssymb}
\usepackage{graphicx}
\usepackage{esint}
\usepackage[unicode=true,pdfusetitle,
 bookmarks=true,bookmarksnumbered=false,bookmarksopen=false,
 breaklinks=false,pdfborder={0 0 1},backref=false,colorlinks=false]
 {hyperref}

\makeatletter

\providecommand{\tabularnewline}{\\}

\numberwithin{equation}{section}
\numberwithin{figure}{section}
\newcommand{\lyxaddress}[1]{
\par {\raggedright #1
\vspace{1.4em}
\noindent\par}
}

\makeatother

\begin{document}

\title{Inflation in a two 3-form fields scenario}

\author{$^{\left(1,\,2\right)}$K. Sravan Kumar%
\thanks{sravan@ubi.pt%
} , $^{\left(1,\,2\right)}$Jo\~ao Marto%
\thanks{jmarto@ubi.pt%
} ,\\
$^{\left(3\right)}$Nelson J. Nunes%
\thanks{njnunes@fc.ul.pt%
} , $^{\left(1,\,2\right)}$Paulo Vargas Moniz%
\thanks{pmoniz@ubi.pt%
}}

\maketitle

\lyxaddress{$^{\left(1\right)}${\small{Departamento de F\'{i}sica, Universidade
da Beira Interior, 6200 Covilh\~a, Portugal.}}}

\lyxaddress{$^{\left(2\right)}${\small{Centro de Matem\'atica e Aplica\c{c}\~oes
da Universidade da Beira Interior (CMA-UBI).}}}

\lyxaddress{$^{\left(3\right)}${\small{Faculty of Sciences and Centre for Astronomy
and Astrophysics, University of Lisbon, 1749-016 Lisbon, Portugal.}}}
\begin{abstract}
A setting constituted by $\mathbb{N}$ 3-form fields, without any
direct interaction between them, minimally coupled to gravity, is
introduced in this paper as a framework to study the early evolution
of the universe. We focus particularly on the two 3-forms case. An
inflationary scenario is found, emerging from the coupling to gravity.
More concretely, the fields coupled in this manner exhibit a complex
interaction, mediated by the time derivative of the Hubble parameter.
Our investigation is supported by means of a suitable choice of potentials,
employing numerical methods and analytical approximations. In more
detail, the oscillations on the small field limit become correlated,
and one field is intertwined with the other.  In this type of solution,
a varying sound speed is present, together with the generation of
isocurvature perturbations. The mentioned features allow to consider
an interesting model, to test against observation. It is subsequently
shown how our results are consistent with current CMB data (viz.Planck
and BICEP2).
\end{abstract}

\section{Introduction}

The recent cosmic microwave background (CMB) studies, developed in
the context of the Planck mission \cite{PconInflation,Ade:2013lta,Ade:2013mta,Bennett:2012fp},
have brought important constraints into the domain of viable inflationary
scenarios. In particular, these accurate measurements seem to favor
simple inflationary slow roll models. They confirm with high accuracy
that the universe is almost spatially flat; the fluctuations are nearly
Gaussian, adiabatic, and showing a nearly scale invariant power spectrum.
The observations also give strong support for a $\Lambda$CDM Universe.
An inflationary period driven by a single scalar field has been the
most widely studied approach in the literature. Nevertheless, our
lack of a precise knowledge regarding the connection between the early
universe and the standard model of particle physics allow us to consider
models of inflation driven by multiple fields \cite{David-wands}.
Moreover, recent advances in experimental particle physics, such as
the evidence towards the discovery of the Higgs boson at a much lower
mass scale than expected, motivate the study of multiple inflaton
field scenarios. Such attempts have been considered in the past and
the literature contains several examples, namely hybrid inflation
\cite{hybrid}, assisted inflation \cite{Liddle-1,Copeland:1999cs}
and k-inflation \cite{Ohashi:2011na,Arroja:2008yy} models. In the
context of the string compactification we may also refer the multiple
axion and assisted chaotic models \cite{dimopoulos,Kanti:1999ie}.
Therefore, it is important to test inflationary models against Planck
data. 

Besides scalar field models, inflation driven by higher spinor fields
has also been investigated. These models have a crucial importance
due to the possible connection to string compactification scenarios\cite{Frey:2002qc,Gubser:2000vg}.
Among these approaches, we mention inflation obtained by means of
vector fields \cite{F89,KM08,MV08}. Other important examples are
$p$-form inflationary models \cite{GK09,KMP09}. Therein, the scalar
and vector fields constitute%
\footnote{It is shown that higher $p$-form fields can ultimately be dualized
to 0 and 1-forms.%
} the 0-form and 1-form cases in the wider classification of $p$-forms.
A most relevant feature of non minimal interactions of massive $p$-forms
with gravity is that it clearly supports an inflationary background
\cite{KMP09,kobayoko}. However, it has been shown that 1-form and
2-form are plagued by a ghost instability during inflation and are
also unable to provide a satisfactory explanation of the CMB isotropy.
Nevertheless, 0-form and 3-form fields are compatible with homogeneity
and isotropy. Ref. \cite{kobayoko} includes a study about p-form
inflation. It is concluded that 3-form field inflation is similar
to the one driven with a scalar field and that it is free from negative
values for the squared propagation speed of the gravitational waves
(which are characteristic of large field 2-form models). Considered
as a suitable alternative to the traditional scalar field inflation,
single 3-form inflation has been introduced and studied in Ref. \cite{Nunes-1,Nunes-2,Felice-1}.
In \cite{Felice-1} a suitable choice of the potential for the 3-form
has been proposed in order to avoid ghosts and Laplacian instabilities;
the authors have shown that potentials showing a quadratic dominance,
in the small field limit, would introduce sufficient oscillations
for reheating \cite{felice2} and would be free of ghost instabilities.

In this work, we extend the single 3-form framework to $\mathbb{N}$
3-forms, to investigate in particular the early universe driven by
two 3-forms. Being more precise, in section \ref{N-form} we identify
basic features of $\mathbb{N}$ 3-forms slow roll solutions, which
can be classified into two types. We also discuss how the inflaton
mass can be brought to lower energy scales, for large values of $\mathbb{N}$.
In section \ref{two 3-form} we examine the possible inflationary
solutions, when two 3-forms are present. There are two classes, namely
solutions not able to generate isocurvature perturbations (type I);
and solutions with inducing isocurvature effects (type II). Our major
objective is to understand and explore the cosmological consequences
of type II solutions. More concretely, we will show that type II brings
about interesting effects. In particular, we use a dynamical system
analysis and conclude that type I case is very similar to single 3-form
inflation \cite{Felice-1,Nunes-2}. Type II case, however, characterizes
a new behavior, through curved trajectories in field space. Moreover,
type II inflation is clearly dominated by the gravity mediated coupling
term which appears in the equations of motion. We present and discuss
type II solutions for several classes of potentials, which are free
from ghost instabilities \cite{Felice-1} and show evidence of a satisfactory
oscillatory behavior at the end of the two 3-forms driven inflation
period. We also provide analytical arguments to explain the oscillatory
behavior present in this choice of potentials. In addition, we calculate
the speed of sound, $c_{s}^{2}$, of adiabatic perturbations for two
3-forms and show it has significant variations during inflation for
type II solutions. In section \ref{powerspectra} we discussed adiabatic
and entropy perturbations for two 3-form fields, using a dualized
action \cite{NunesNG,Langlois:2008mn}. We distinguish, type I and
type II solutions with respect to isocurvature perturbations and calculate
the power spectrum expression \cite{lyth2009primordial}. In section
\ref{sec:Planck} we present how our inflationary setting can fit
the tensor scalar ratio, spectral index and its running provided by
the Planck data as well as the recent BICEP2 results \cite{PconInflation,Ade:2014xna}.
In section \ref{discussion} we summarize and discuss the model herein
investigated and propose subsequent lines to explore. Appendix \ref{appendix}
provides stability analysis regarding type I solutions.

\section{$\mathbb{N}$ 3-form fields model}

\label{N-form}In this section, we generalize the background equations
associated to a single 3-form field, which has been studied in \cite{Nunes-1,Nunes-2,Felice-1},
to $\mathbb{N}$ 3-form fields. We take a flat Friedmann-Lemaitre-Robertson-Walker
(FLRW) cosmology, described with the metric 
\begin{equation}
ds^{2}=-dt{}^{2}+a^{2}(t)d\boldsymbol{x}^{2},\label{FRW-metric1-1}
\end{equation}
where $a(t)$ is the scale factor with $t$ being the cosmic time.
The general action for Einstein gravity and\textit{\emph{ $\mathbb{N}$}}
3-form fields is written as 
\begin{equation}
S=-\int d^{4}x\sqrt{-g}\left[\frac{1}{2\kappa^{2}}R-\sum_{n=1}^{\mathbb{N}}\left(\frac{1}{48}F_{n}^{2}+V_{n}(A_{n}^{2})\right)\right]\,,\quad\kappa^{2}=8\pi G\,,\label{N3F-action-1}
\end{equation}
where $A_{\beta\gamma\delta}^{(n)}$ is the $n$th 3-form field and
we have squared the quantities by contracting all the indices. The
strength tensor of the 3-form is given by%
\footnote{Throughout this paper, the Latin index $n$ will be used to refer
the number of the quantity (or the 3-form field) or the $n$th quantity/field.
The other Latin indices, which take the values $i,j=1,2,3$, will
indicate the three dimensional quantities; whereas the Greek indices
will be used to denote four-dimensional quantities and they stand
for $\mu,\nu=0,1,2,3$.%
} 
\begin{equation}
F_{\alpha\beta\gamma\delta}^{(n)}\equiv4\nabla_{[\alpha}A_{\beta\gamma\delta]}^{(n)},\label{N3f-Maxw-1}
\end{equation}
where antisymmetrization is denoted by square brackets. As we have
assumed a homogeneous and isotropic universe, the 3-form fields depend
only on time and hence only the space like components will be dynamical,
thus their nonzero components are given by 
\begin{equation}
A_{ijk}^{(n)}=a^{3}(t)\epsilon_{ijk}\chi_{n}(t)\quad\Rightarrow A_{n}^{2}=6\chi_{n}^{2},\label{NNZ-comp-1-1}
\end{equation}
where $\chi_{n}(t)$ is a comoving field associated to the $n$th
3-form field and $\epsilon_{ijk}$ is the standard three dimensional
Levi-Civita symbol. Also note that by introducing the more convenient
field $\chi_{n}(t)$, which is related to the corresponding 3-form
field by the above relation, we have, subsequently, the following
system of equations of motion for $\mathbb{N}$ 3-form fields 
\begin{eqnarray}
\ddot{\chi}_{n}+3H\dot{\chi}_{n}+3\dot{H}\chi_{n}+V_{(n),\chi_{n}} & =0,\label{NDiff-syst-1-1}
\end{eqnarray}
where $V_{(n)}\equiv V_{n}\left(\chi_{n}\right)$ and $V_{(n),\chi_{n}}\equiv\frac{dV_{(n)}}{d\chi_{n}}$.
For each value of $n$ , each of the Eqs. (\ref{NDiff-syst-1-1})
are not independent: it is straightforward to see that a peculiar
coupling is present through the Hubble parameter derivative, $\dot{H}$.
This fact will play a crucial role, establishing different classes
of inflationary behavior when more than one three-from field is employed.
In this setting, the gravitational sector equations are given by 
\begin{equation}
\begin{aligned}H^{2} & =\frac{\kappa^{2}}{3}\left\{ \frac{1}{2}\sum_{n=1}^{\mathbb{N}}\left[\left(\dot{\chi}_{n}+3H\chi_{n}\right){}^{2}+2V_{(n)}\right]\right\} ,\end{aligned}
\label{NFriedm-1-1}
\end{equation}
\begin{equation}
\dot{H}=-\frac{\kappa^{2}}{2}\left[\sum_{n=1}^{\mathbb{N}}V_{(n),\chi_{n}}\chi_{n}\right].\label{NHdot-1-1}
\end{equation}
Therefore, the mentioned (gravity mediated) coupling between the several
$\mathbb{N}$ 3-form fields will act through the gravitational sector
of the equations of motion. The total energy density and pressure
of the $\mathbb{N}$ 3-form fields read 
\begin{equation}
\rho_{\mathbb{N}}=\frac{1}{2}\sum_{n=1}^{\mathbb{N}}\left[\left(\dot{\chi}_{n}+3H\chi_{n}\right){}^{2}+2V_{(n)}\right],\label{Ndens-1-1}
\end{equation}
\begin{equation}
p_{\mathbb{N}}=-\frac{1}{2}\sum_{n=1}^{\mathbb{N}}\left[\left(\dot{\chi}_{n}+3H\chi_{n}\right){}^{2}+2V_{(n)}-2V_{(n),\chi_{n}}\chi_{n}\right].\label{Npress-1-1}
\end{equation}
We rewrite Eq. (\ref{NDiff-syst-1-1}) as 
\begin{eqnarray}
\ddot{\chi}_{n}+3H\dot{\chi}_{n}+V_{n,\chi_{n}}^{\textrm{eff}} & =0,,\label{NDiff-syst-1-1-2}
\end{eqnarray}
where 
\begin{equation}
V_{n,\chi_{n}}^{\textrm{eff}}\equiv3\dot{H}\chi_{n}+V_{(n),\chi_{n}}=V_{(n),\chi_{n}}\left[1-\frac{3\kappa^{2}}{2}\chi_{n}^{2}\right]-\frac{3}{2}\kappa^{2}\chi_{n}\left[\overset{\mathbb{N}}{\underset{\underset{n\neq m}{m=1}}{\sum}}V_{(m),\chi_{m}}\chi_{m}\right].\label{NGen-X-Pot-1}
\end{equation}
In order to describe the dynamics of the 3-form fields, we express
the equations of motion in terms of the dimensionless variables, 
\begin{eqnarray}
x_{n} & \equiv\!\!\! & \kappa\,\chi_{n},\nonumber \\
w_{n} & \equiv & \frac{\chi'_{n}+3\chi_{n}}{\sqrt{6}},\label{NQuad-omega-1}
\end{eqnarray}
where $x'_{n}\equiv dx_{n}/dN$ in which the number of e-folds of
inflationary expansion is $N=\ln\, a(t)$. Thus, we get 
\begin{eqnarray}
H^{2}x_{n}''+\left(3H^{2}+\dot{H}\right)x_{n}'+V_{n,x_{n}}^{\textrm{eff}} & = & 0.\label{NQuad-X-diff-1}
\end{eqnarray}
The Friedmann constraint is written as 
\begin{equation}
H^{2}=\frac{1}{3}\frac{\sum_{n=1}^{\mathbb{N}}V_{(n)}}{\left(1-w^{2}\right)},\label{NQuad-Fried-1}
\end{equation}
where 
\[
w^{2}\equiv\sum_{n=1}^{\mathbb{N}}w_{n}^{2}.
\]
Employing the dimensionless variables (\ref{NQuad-omega-1}), the
equations of motion (\ref{NQuad-X-diff-1}) can be rewritten in the
autonomous form as 
\begin{equation}
x'_{n}=3\left(\sqrt{\frac{2}{3}}w_{n}-x_{n}\right),\label{Nauton-n-n}
\end{equation}
\begin{eqnarray}
w'_{n} & = & \frac{3}{2}\frac{V_{(n),x_{n}}}{V}\left(1-w^{2}\right)\left(x_{n}w_{n}-\sqrt{\frac{2}{3}}\right)+\frac{3}{2}\left(1-w^{2}\right)\frac{1}{V}w_{n}\overset{\mathbb{N}}{\underset{\underset{n\neq m}{m=1}}{\sum}}x_{m}V_{(m),x_{m}},\label{Nwprime-n}
\end{eqnarray}
where 
\begin{equation}
V=\sum_{n=1}^{\mathbb{N}}V_{(n)}.
\end{equation}

\subsection{Initial conditions and slow roll inflation }

\label{sub:Initial cond.}

Analogous to the scalar field \cite{Liddle-1} as well as single 3-form
\cite{Nunes-1,Felice-1} inflationary models, the so-called slow roll
parameters are taken as $\epsilon\equiv-\dot{H}/H^{2}=-d\ln H/dN$
and $\eta\equiv\epsilon'/\epsilon-2\epsilon$, which, for our model,
are given by%
\footnote{Equivalently solely in terms of $x_{n}$and $w_{n}$,$\eta=\frac{\sum_{n=1}^{\mathbb{N}}3\left(\sqrt{\frac{2}{3}}w_{n}-x_{n}\right)\left(V_{n,x_{n}}+V_{n,x_{n}x_{n}}x_{n}\right)}{\sum_{n=1}^{\mathbb{N}}V_{n,x_{n}}x_{n}}$%
}

\begin{equation}
\epsilon=\frac{3}{2}\frac{\sum_{n=1}^{\mathbb{N}}V_{n,x_{n}}x_{n}}{V}\left(1-w^{2}\right),\label{N-3F-epsilon-1}
\end{equation}

\begin{equation}
\eta=\frac{\sum_{n=1}^{\mathbb{N}}x'_{n}\left(V_{n,x_{n}}+V_{n,x_{n}x_{n}}x_{n}\right)}{\sum_{n=1}^{\mathbb{N}}V_{n,x_{n}}x_{n}}.\label{N-3F-eta-1}
\end{equation}
We can see from (\ref{N-3F-epsilon-1}) and (\ref{N-3F-eta-1}) that,
for $\mathbb{N}$ 3-form fields, one manner to establish a sufficient
condition for inflation (with the slow roll parameters $\epsilon\ll1$
and $\eta\ll1$) is by means of 
\begin{equation}
\begin{cases}
1-\sum_{n=1}^{\mathbb{N}}w_{n}^{2} & \approx0,\\
x'_{n} & \approx0
\end{cases}\label{Ident-sl-cond}
\end{equation}
It is important, however, to also consider another (albeit less obvious)
possibility, which is to have instead 
\begin{equation}
\begin{cases}
1-\sum_{n=1}^{\mathbb{N}}w_{n}^{2}\approx0,\\
\sum_{n=1}^{\mathbb{N}}x'_{n}\left(V_{n,x_{n}}+V_{n,x_{n}x_{n}}x_{n}\right) & \approx0.
\end{cases}\label{NonIdent-sl-cond}
\end{equation}
The condition expressed in (\ref{NonIdent-sl-cond}) means that the
inclusion of more than one 3-form field allows the emergence of an
inflationary scenario without even requiring that $x_{n}'\approx0$.
Therefore, we can expect to have different behaviors, in contrast
to the ones usually found in models with just one 3-form. The different
$\mathbb{N}$ 3-form fields will evolve in an intricate correlated
way in order to satisfy (\ref{NonIdent-sl-cond}). This possibility
will deserve a more detailed analysis in the next sections. We should
note that all the derived equations in this section reduce to the
single one 3-form case when $\mathbb{N}=1$, as expected.

\subsection{Inflaton mass}

\label{low3formmass}Returning to the condition (\ref{Ident-sl-cond}),
we have

\[
\sum_{n=1}^{\mathbb{N}}x_{n}^{2}\thickapprox\,\frac{2}{3\mathbb{N}}.
\]
Note that the Friedman constraint (\ref{NQuad-Fried-1}) does not
hold precisely at $\sum_{n=1}^{\mathbb{N}}x_{n}^{2}=\,2/3\mathbb{N}$,
and $x'_{n}=0$. If we assume a symmetric situation, where all $w_{n}$
are equal during inflation, i.e, if all fields come to the same value
during inflation, then $x_{n}\left(N\right)$ will take a constant
value 
\begin{equation}
x_{p}=\,\sqrt{\frac{2}{3\mathbb{N}}}.\label{N3formsymplateau}
\end{equation}
In this symmetric situation, all the 3-form fields will behave identically
during inflation. If $\mathbb{N}$ is very large, the plateau in $x_{n}\left(\textrm{N}\right)$
converges towards zero ($x_{p}=\sqrt{\frac{2}{3\mathbb{N}}}\rightarrow0$
as $\mathbb{N\rightarrow\infty}$, for the symmetric case where all
$w_{n}$ are equal). The initial conditions for the one single 3-form
inflation case were discussed in \cite{Felice-1}. The reduction of
the plateau energy scale for $\mathbb{N}$ 3-forms can have a nontrivial
consequence, which is to bring the inflaton mass well below Planck
mass. This is illustrated by the following analysis. Expressing the
scale of the $\mathbb{N}$ 3-forms plateau (\ref{N3formsymplateau})
in terms%
\footnote{An equation similar to (\ref{inflaton mass}) was also proposed to
study the case of $\mathbb{N}$-flation (multiple axion inflation)
model in \cite{dimopoulos}. Whereas in M-flation (Matrix inflation)
there will be additional corrections, depending on $\mathbb{N}$,
involved in Planck mass and arising from standard loop effects \cite{Ashoorioon:2011ki,Ashoorioon:2011aa}. %
} of the reduced Planck mass $\left(M_{{\rm Pl}}=1/\sqrt{8\pi G}\right)$,
\begin{equation}
\chi_{p}\thickapprox\sqrt{\frac{2}{3\mathbb{N}}}M_{{\rm Pl}},\label{inflaton mass}
\end{equation}
let us assume that all the 3-form fields behave in the same way, reaching
a constant value $x_{p}$ during inflation and starting to oscillate
by the end of inflation. Subsequently, we rewrite the Friedmann constraint
(\ref{NQuad-Fried-1}) for this case as,

\begin{equation}
H^{2}=\frac{1}{3}\frac{\widetilde{V}}{\left(1-w^{2}\right)},\label{N3form mass}
\end{equation}
with $\widetilde{V}=\underset{n}{\sum}\, V_{(n)}$ , $V_{(n)}=V_{0n}\, f_{n}\left(x_{n}\right)$
, and where $f_{n}\left(x_{n}\right)$ are dimensionless functions.
Comparing the above Eq. (\ref{N3form mass}) with the Friedmann constraint
of a single 3-form field case, we get

\begin{equation}
\widetilde{V}=V_{1},\label{Mass label}
\end{equation}
where $V_{1}=V_{0}\, f_{1}(x_{1})$ is the potential for the single
3-form field case. If we choose $V_{01}=V_{02}=\cdots=V_{0n}=V_{0\mathbb{N}}$,
which means that the energy scales of the potentials are the same,
and also assume that $x_{n}=x_{p}$ for all $n$ in Eq. (\ref{Mass label}),
we get 
\begin{equation}
\frac{V_{0\mathbb{N}}}{V_{0}}=\frac{f_{1}\left(x_{1}\right)}{f_{n}\left(x_{n}\right)}\label{energy lable}
\end{equation}
Let us consider the power law potential $f=x^{l}$ , for a 3-form.
If we substitute the corresponding value of the plateau for $\mathbb{N}$
3-form $\left(x_{p}=\sqrt{\frac{2}{3\mathbb{N}}}\right)$ and of the
single 3-form case $\left(x_{1p}=\sqrt{\frac{2}{3}}\right)$ in Eq.
(\ref{energy lable}), then we can have the following ratio of energy
scales for the potentials, of $\mathbb{N}$ 3-forms and single 3-form

\begin{equation}
\frac{V_{0\mathbb{N}}}{V_{0}}=\mathbb{\mathbb{N}}^{1-\frac{l}{2}}.\label{energy scale}
\end{equation}
We can translate this argument in terms of the inflaton mass, which
is defined to be the square root of the second derivative of the potential.
Therefore, the ratio between the inflaton masses corresponding to
the $\mathbb{N}$ 3-forms potential $\left(m_{\mathbb{N}}\right)$,
and the single 3-form $\left(m_{1}\right)$, for a power law potential
$\left(x^{l}\right)$, is given by

\begin{equation}
\frac{m_{\mathbb{N}}}{m_{1}}\equiv\sqrt{\frac{V_{0\mathbb{N}}\, x_{p}^{l-2}}{V_{0}\, x_{1p}^{l-2}}}=\dfrac{1}{\mathbb{\mathbb{N}}^{\frac{l}{2}-1}}\label{massscale}
\end{equation}

In the case of quadratic potentials $\left(l=2\right)$, there is
no reduction in the inflaton mass, similar to the single field case.
This fact was expected since the 3-form dual action \cite{NunesNG},
when a quadratic potential is included, can be shown to be equivalent
to a canonical scalar field inflation. This dual action picture enhance
the fact that the 3-form inflation can encompass both features of
canonical and non canonical models. In the case where $l>2$ in (\ref{massscale}),
it is possible to bring down the mass of the inflaton to lower energy
scales by increasing the number of 3-form fields. This property opens
the possibility to make a connection with particle physics since the
Higgs mass was already referenced at a 125 GeV scale.


\section{Two 3-form fields model}

\label{two 3-form}In the present section, we would like to concentrate
on the case where only two 3-form fields are present %
\footnote{A generalization from two to $\mathbb{N}$ 3-form fields is left for
a future work, although, whenever possible, we will point out the
main aspects that can straightforwardly be generalized.%
}. Accordingly, we will rewrite some of the equations as follows. Thus,
the nonzero components of Eq.~(\ref{NNZ-comp-1-1}) are 
\begin{equation}
\begin{aligned}A_{ijk}^{\left(1\right)}=a^{3}(t)\epsilon_{ijk}X_{1}(t),\hspace{1cm}A_{ijk}^{\left(2\right)}=a^{3}(t)\epsilon_{ijk}X_{2}(t),\\
\Rightarrow A_{\left(1\right)}^{2}=6X_{\left(1\right)}^{2},\hspace{2cm}A_{\left(2\right)}^{2}=6X_{\left(2\right)}^{2}.
\end{aligned}
\label{2NZ-comp-1}
\end{equation}
Also, we rewrite equations of motion (\ref{NQuad-X-diff-1}) in terms
of our dimensionless variables as 
\begin{eqnarray}
H^{2}x''_{1}+\left(3H^{2}+\dot{H}\right)x'_{1}+V_{1,x_{1}}^{\textrm{eff}} & = & 0,\label{2Quad-X-diff}\\
H^{2}x''_{2}+\left(3H^{2}+\dot{H}\right)x'_{2}+V_{2,x_{2}}^{\textrm{eff}} & = & 0,\label{2Quad-Y-diff}
\end{eqnarray}
where the Friedmann and acceleration equations are given by 
\begin{equation}
\begin{aligned}H^{2} & =\frac{1}{3}\frac{V_{1}(x_{1})+V_{2}(x_{2})}{\left(1-w_{1}^{2}-w_{2}^{2}\right)},\end{aligned}
\label{2Quad-Fried}
\end{equation}
\begin{equation}
\dot{H}=-\frac{1}{2}\left(V_{1,x_{1}}x_{1}+V_{2,x_{2}}x_{2}\right).\label{2Hdot-1}
\end{equation}

In order to further discuss suitable initial conditions, the slow
roll conditions $\epsilon,\left|\eta\right|\ll1$ suggests the equation
of a circle (of unit radius), as 
\begin{equation}
w_{1}^{2}+w_{2}^{2}\approx1,\label{eps -constr-1}
\end{equation}
which we rewrite in terms of trivial parametric relations as 
\begin{equation}
\begin{aligned}w_{1}\approx\:\cos\theta\,,\\
w_{2}\approx\:\sin\theta\,.
\end{aligned}
\label{a-value}
\end{equation}
Subsequently, from (\ref{Nauton-n-n}), we can establish the initial
conditions for the field derivatives 
\begin{eqnarray}
\begin{cases}
x'_{1}\approx3\left(\sqrt{\frac{2}{3}}\:\cos\theta-x_{1}\right),\\
x'_{2}\approx3\left(\sqrt{\frac{2}{3}}\;\sin\theta-x_{2}\right).
\end{cases}\label{x1x2plateau}
\end{eqnarray}
Since Eq.~(\ref{a-value}) can be satisfied by assigning many different
continuous values of the new parameter $\theta$, we, therefore, anticipate
to investigate diverse solutions. More precisely, a particular choice
of this parameter will affect the way (\ref{N-3F-eta-1}), ( i.e,
the value of $\eta$) will depend on the two 3-form fields. Before
proceeding, let us mention that for two 3-forms inflation, we choose
herein initial conditions for the fields above or below the value
given by (\ref{N3formsymplateau}), which are expected to influence
the number of e-foldings. In particular, we will investigate the asymmetric
situation, when each $w_{n}$ is different%
\footnote{For example, if we take $w_{1}=1$ and $w_{n>1}=0$, we then find
a scenario similar to one single 3-form field driving the inflation
and where all the other fields approach zero.%
}, which will provide a new behavior with respect to inflation.


\subsection{Type I inflation ($x_{n}^{\prime}\approx0$)}

\label{sub: Type I}As we have established in section \ref{sub:Initial cond.},
the slow roll conditions enable us to find two types of inflationary
solutions, according to relation (\ref{Ident-sl-cond})-(\ref{NonIdent-sl-cond}).
In this subsection and the following, we investigate them in more
detail. In type I solution, presented in this subsection, the 3-form
fields which are responsible for driving the inflationary period,
will be displaying $x'_{n}\approx0$. The following is a stability
analysis for this type, presented in a dynamical system context. Whenever
necessary, we will complement this study by a numerical discussion.

Let us remind the autonomous system of equations for the field $x_{1}$,
\begin{align}
x'_{1}= & 3\left(\sqrt{\frac{2}{3}}w_{1}-x_{1}\right),\label{Dyn-x1}\\
w'_{1}= & \frac{3}{2}\left(1-\left(w_{1}^{2}+w_{2}^{2}\right)\right)\left(\lambda_{1}\left(x_{1}w_{1}-\sqrt{\frac{2}{3}}\right)+\lambda_{2}x_{2}w_{1}\right),\label{Dyn-w1}
\end{align}
and also for $x_{2}$, 
\begin{align}
x'_{2}= & 3\left(\sqrt{\frac{2}{3}}w_{2}-x_{2}\right),\label{Dyn-x2}\\
w'_{2}= & \frac{3}{2}\left(1-\left(w_{1}^{2}+w_{2}^{2}\right)\right)\left(\lambda_{2}\left(x_{2}w_{2}-\sqrt{\frac{2}{3}}\right)+\lambda_{1}x_{1}w_{2}\right),\label{Dyn-w2}
\end{align}
where $\lambda_{n}=V_{\left(n\right),x_{n}}/V$. Notice that, Eqs.~(\ref{Dyn-x1})-(\ref{Dyn-w1})
are coupled with Eqs.~(\ref{Dyn-x2})-(\ref{Dyn-w2}). With the variables
$(x_{n},w_{n})$, let $f_{1}:=dx_{1}/dN$, $f_{2}:=dw_{1}/dN$, $f_{3}:=dx_{2}/dN$
and $f_{4}:=dw_{2}/dN$. The critical points are located at the field
space coordinates $\left(x_{c}\right)$ and are obtained by setting
the condition $(f{}_{1},\, f_{2},\, f_{3},\, f_{4})|_{x_{{\rm c}}}=0$.

To determine the stability of the critical points, we need to perform
linear perturbations around each of them by using $x(t)=x_{{\rm c}}+\delta x(t)$;
this results in the equations of motion $\delta x'={\cal M}\delta x$,
where ${\cal M}$ is the Jacobi matrix of each critical point whose
components are ${\cal M}_{ij}=(\partial f_{i}/\partial x_{j})|_{x_{{\rm c}}}$.
A critical point is called stable (unstable) whenever the eigenvalues
$\zeta_{i}$ of ${\cal M}$ are such that $\mathsf{Re}(\zeta_{i})<0$
($\mathsf{Re}(\zeta_{i})>0$) \cite{Khalil}. If $\mathsf{Re}(\zeta_{i})=0$,
then other methods should be employed to further assess the stability
of the critical point. Among different approaches, we have the center
manifold theorem \cite{Carr,Guckenheimer,Khalil,Lazkoz-1} or, alternatively,
we can consider a perturbative expansion to nonlinear order as in
Refs. \cite{FKW12,Nunes-2}. In this work we will follow the last
mentioned method, whenever necessary.

The autonomous dynamical Eqs. (\ref{Dyn-x1})-(\ref{Dyn-w2}) fixed
points are given by 
\begin{equation}
\begin{aligned}x_{1c}= & \sqrt{\frac{2}{3}}w_{1},\hspace{1cm}w_{1c}=\sqrt{\frac{2}{3}}\frac{\lambda_{1}}{\lambda_{1}x_{1}+\lambda_{2}x_{2}},\\
x_{2c}= & \sqrt{\frac{2}{3}}w_{2},\hspace{1cm}w_{2c}=\sqrt{\frac{2}{3}}\frac{\lambda_{2}}{\lambda_{1}x_{1}+\lambda_{2}x_{2}}.
\end{aligned}
\label{fixed-P1}
\end{equation}
If $\lambda_{1}\neq0$ and $\lambda_{2}\neq0$ (otherwise, $V_{1,x_{1}}=0$
and $V_{2,x_{2}}=0$ ), Eqs. (\ref{fixed-P1}) can be rewritten as
\begin{equation}
\begin{aligned}x_{1c}= & \sqrt{\frac{2}{3}}w_{1},\hspace{1cm}w_{1c}=\frac{\lambda_{1}}{\sqrt{\lambda_{1}^{2}+\lambda_{2}^{2}}},\\
x_{2c}= & \sqrt{\frac{2}{3}}w_{2},\hspace{1cm}w_{2c}=\frac{\lambda_{2}}{\sqrt{\lambda_{1}^{2}+\lambda_{2}^{2}}}.
\end{aligned}
\label{fixed-P2}
\end{equation}

Generically, with a inflationary stage as a target, the fixed points
coordinates must satisfy $w_{1c}^{2}+w_{2c}^{2}\simeq1$. This last
condition is required to satisfy the slow roll condition (\ref{Ident-sl-cond}).
Therefore, we can define as well 
\begin{equation}
\begin{aligned}w_{1c}=\:\cos\theta\,,\\
w_{2c}=\:\sin\theta\,.
\end{aligned}
\label{Type I omega}
\end{equation}
Note that, when we consider the field coordinates in (\ref{fixed-P1}),
also $x_{1c}$ and $x_{2c}$ are constrained by $x_{1c}^{2}+x_{2c}^{2}=2/3$.
Consequently, the dynamical system (\ref{Dyn-x1})-(\ref{Dyn-w2})
has fixed points with only two independent degrees of freedom, which
can be chosen to be the pair $\left(x_{1c},\, w_{1c}\right)$. Therefore,
the critical points or inflationary attractors are found by solving
the following expression $w_{1c}^{2}+w_{2c}^{2}=1$, which upon substitution
gives, 
\begin{equation}
\left(\sqrt{\frac{2}{3}}\frac{\lambda_{1}}{\lambda_{1}x_{1c}+\lambda_{2}x_{2c}}\right)^{2}+\left(\sqrt{\frac{2}{3}}\frac{\lambda_{2}}{\lambda_{1}x_{1c}+\lambda_{2}x_{2c}}\right)^{2}=1.\label{theta label}
\end{equation}
It is clear from the Eq. (\ref{theta label}) that the location of
critical points depends on the choice of 3-form potentials. For example
let us take $V_{1}=x_{1}^{n}$ and $V_{2}=x_{2}^{m}$. It follows
that\textcolor{blue}{{} } 
\begin{equation}
\left(\frac{n\left(\sqrt{\frac{2}{3}}\right)^{n}\left(\cos\theta\right)^{n-1}}{n\left(\sqrt{\frac{2}{3}}\cos\theta\right)^{n}+m\left(\sqrt{\frac{2}{3}}\sin\theta\right)^{m}}\right)^{2}+\left(\frac{m\left(\sqrt{\frac{2}{3}}\right)^{m}\left(\sin\theta\right)^{m-1}}{n\left(\sqrt{\frac{2}{3}}\cos\theta\right)^{n}+m\left(\sqrt{\frac{2}{3}}\sin\theta\right)^{m}}\right)^{2}=1.\label{cond-V1-V2}
\end{equation}
From Eq. (\ref{cond-V1-V2}), $\theta=0$ and $\theta=\pi/2$ can
be called as trivial fixed points independent of the choice of $n,m$.
Satisfying the condition (\ref{cond-V1-V2}), for our particular choice
of potentials, allows us to also identify non trivial fixed points
in the range $0<\theta<\pi/2$. To easily identify these, we can extract
a simple constraint from Eq. (\ref{fixed-P2}), given by

\begin{equation}
x_{1c}/x_{2c}=\lambda_{1}/\lambda_{2}\label{simpelconditiontype 1}
\end{equation}
Condition (\ref{simpelconditiontype 1}) is fully consistent with
Eq.(\ref{cond-V1-V2}), except for the trivial fixed points $\theta=0$
and $\theta=\pi/2$. Let us apply the example where $V_{1}=x_{1}^{n}$
and $V_{2}=x_{2}^{m}$, and substituting in the Eq. (\ref{simpelconditiontype 1}),
We have 
\begin{equation}
\frac{n\left(\sqrt{\frac{2}{3}}\cos\theta\right)^{n-2}}{m\left(\sqrt{\frac{2}{3}}\sin\theta\right)^{m-2}}=1\,.\label{n,m.critical points}
\end{equation}
We can read from Eq. (\ref{n,m.critical points}) that for identical
quadratic potentials, i.e., for $n=m=2$, Eq. \ref{simpelconditiontype 1}
is satisfied for all values of $0<\theta<\pi/2$. Identical quadratic
potentials is the only case where we can have an infinite number of
non trivial fixed points. For any other choice of potentials , i.e.,
for $n\neq m$, there will only be a finite number of non trivial
fixed points.

\begin{figure}[h!]
\centering\includegraphics[height=1.9in]{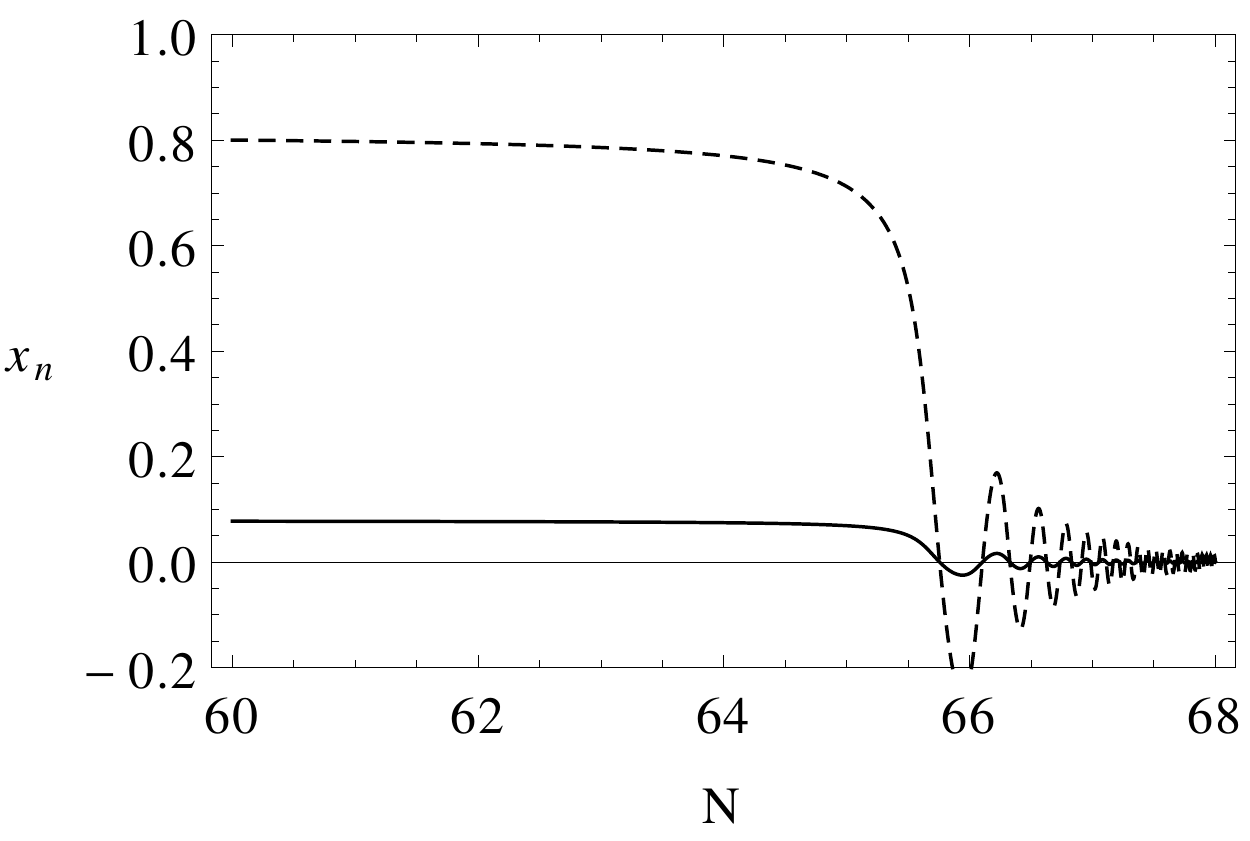}\quad{}\includegraphics[height=1.9in]{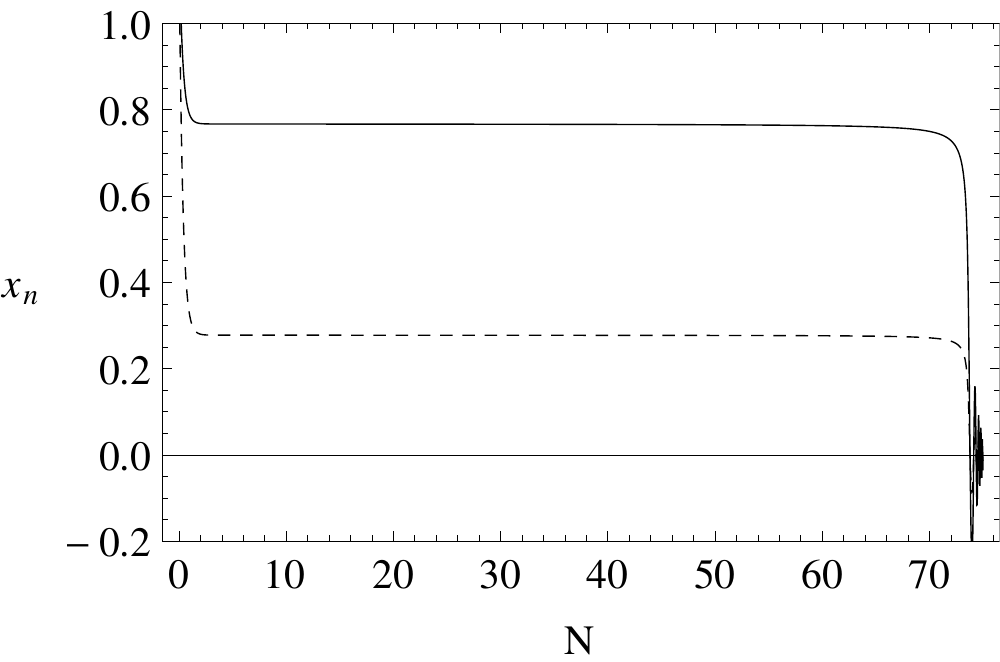}
\caption{Left panel is the graphical representation of the numerical solutions
of (\ref{2Quad-X-diff}) and (\ref{2Quad-Y-diff}) for $x_{1}\left(N\right)$
(full line) and $x_{2}\left(N\right)$ (dashed line) with $\theta\approx\dfrac{\pi}{2}$
for the potentials $V_{1}=x_{1}^{2}$ and $V_{2}=x_{2}^{2}$\textbf{.}
In the right panel, we depict the graphical representation of the
numerical solutions of (\ref{2Quad-X-diff}) and (\ref{2Quad-Y-diff})
for $x_{1}\left(N\right)$ (full line) and $x_{2}\left(N\right)$
(dashed line) with $\theta=\dfrac{\pi}{9}$. We have taken the initial
conditions as $x_{1}(0)=2.1\times\sqrt{\frac{1}{3}}$ and $x_{2}(0)=2.1\times\sqrt{\frac{1}{3}}$.}

\label{fig1} 
\end{figure}

In Fig.~\ref{fig1} we illustrate the evolution of the fields $x_{1}$
and $x_{2}$ for quadratic potentials with $\theta=\pi/2$ and $\theta=\pi/4$.
The asymmetry in choosing $\theta\neq\pi/4$ manifests through one
of the 3-form fields having a plateau slightly higher than the other.

Type I solutions, as far as the stability analysis, are very similar
to the scenario where just one single 3-form field is present. The
novelty here is that we can have solutions as shown in Fig. \ref{fig1}.
Therein, we have a case where we consider that the 3-form fields $x_{1}$
and $x_{2}$ are under the influence of the same kind of quadratic
potential, i.e, $V_{n}=x_{n}^{2}$.

We discuss the stability of these fixed points and their stability
in the Appendix for simple potentials. Other combinations of potentials
can be tested for stability along the same lines presented there.
We summarize the results in Table \ref{potential-stability}. 

\begin{center}
\begin{table}[h!]
\centering%
\begin{tabular}{c|c|c|c|c}
\hline 
{\footnotesize{{}~~$V\left(x_{1}\right)$~~}}  & {\footnotesize{{}~~$V\left(x_{2}\right)$~~}}  & {\footnotesize{{}existence}}  & {\footnotesize{{}stability}}  & {\footnotesize{{}Oscillatory regime}}\tabularnewline
\hline 
\hline 
{\footnotesize{{}$x_{1}^{2}$}}  & {\footnotesize{{}$x_{2}^{2}$}}  & {\footnotesize{{}$0<\theta<\pi/2$}}  & {\footnotesize{{}unstable saddle}}  & {\footnotesize{{}yes}}\tabularnewline
{\footnotesize{{}$x_{1}^{4}+x_{1}^{2}$}}  & {\footnotesize{{}$x_{2}^{4}+x_{2}^{2}$}}  & {\footnotesize{{}$\theta=\left\{ 0,\,\pi/4,\,\pi/2\right\} $}}  & {\footnotesize{{}unstable}}  & {\footnotesize{{}yes}}\tabularnewline
{\footnotesize{{}$x_{1}^{3}+x_{1}^{2}$}}  & {\footnotesize{{}$x_{2}^{3}+x_{2}^{2}$}}  & {\footnotesize{{}$\theta=\left\{ 0,\,\pi/4,\,\pi/2\right\} $}}  & {\footnotesize{{}unstable}}  & {\footnotesize{{}yes}}\tabularnewline
{\footnotesize{{}$\exp\left(x_{1}^{2}\right)-1$}}  & {\footnotesize{{}$\exp\left(x_{2}^{2}\right)-1$}}  & {\footnotesize{{}$\theta=\left\{ 0,\,\pi/4,\,\pi/2\right\} $}}  & {\footnotesize{{}unstable}}  & {\footnotesize{{}yes}}\tabularnewline
{\footnotesize{{}$x_{1}^{2}$}}  & {\footnotesize{{}$x_{2}^{4}+x_{2}^{2}$}}  & {\footnotesize{{} $\theta=\left\{ 0,\,\pi/2\right\} $}}  & {\footnotesize{{}unstable}}  & {\footnotesize{{}yes}}\tabularnewline
{\footnotesize{{}$\exp\left(-x_{1}^{2}\right)$}}  & {\footnotesize{{}$\exp\left(-x_{2}^{2}\right)$}}  & {\footnotesize{{}$\theta=\left\{ 0,\,\pi/4,\,\pi/2\right\} $}}  & {\footnotesize{{}unstable}}  & {\footnotesize{{}no}}\tabularnewline
{\footnotesize{{}$x_{1}^{2}$}}  & {\footnotesize{{}$x_{2}^{4}$}}  & {\footnotesize{{}$\theta=\left\{ 0,\,\pi/3,\,\pi/2\right\} $}}  & {\footnotesize{{}unstable}}  & {\footnotesize{{}yes}}\tabularnewline
{\footnotesize{{}$x_{1}^{n}\quad\left(n>2\right)$}}  & {\footnotesize{{}$x_{2}^{n}\quad\left(n>2\right)$}}  & {\footnotesize{{}$\theta=\left\{ 0,\,\pi/4,\,\pi/2\right\} $}}  & {\footnotesize{{}unstable}}  & {\footnotesize{{}no}}\tabularnewline
\hline 
\hline 
\multicolumn{1}{c}{} & \multicolumn{1}{c}{} & \multicolumn{1}{c}{} & \multicolumn{1}{c|}{} & \tabularnewline
\end{tabular}\caption{Summary of some type I solutions critical points and their properties.}

\label{potential-stability} 
\end{table}

\par\end{center}

\subsection{Type II inflation $\left(x_{n}^{\prime}\not\approx0\right)$ }

\label{sub: Type II} 
\begin{figure}[ht!]
\centering\includegraphics[height=2in]{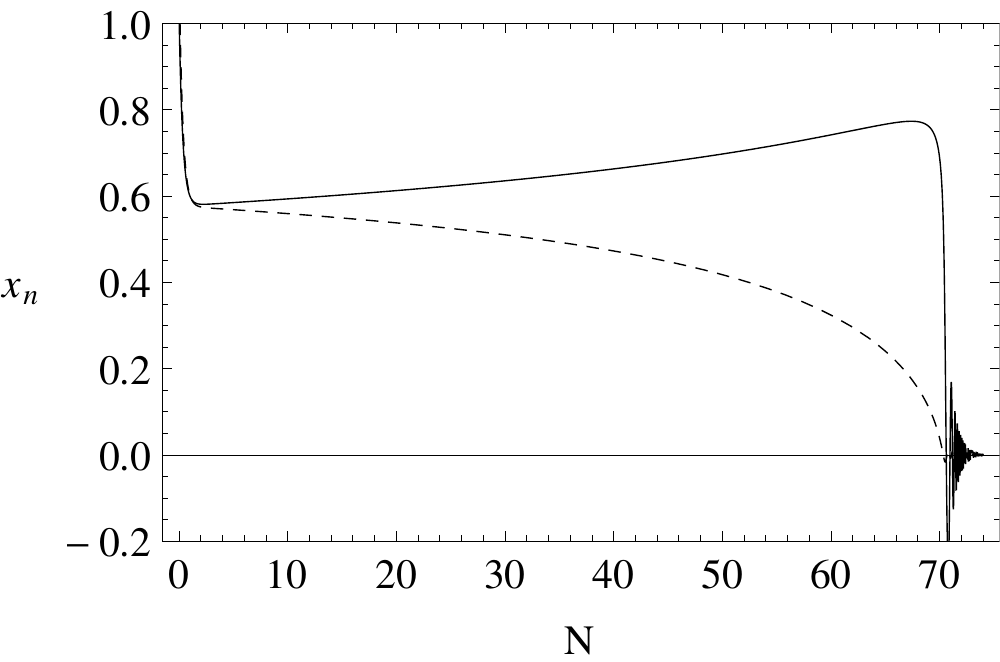}\quad{}\includegraphics[height=2in]{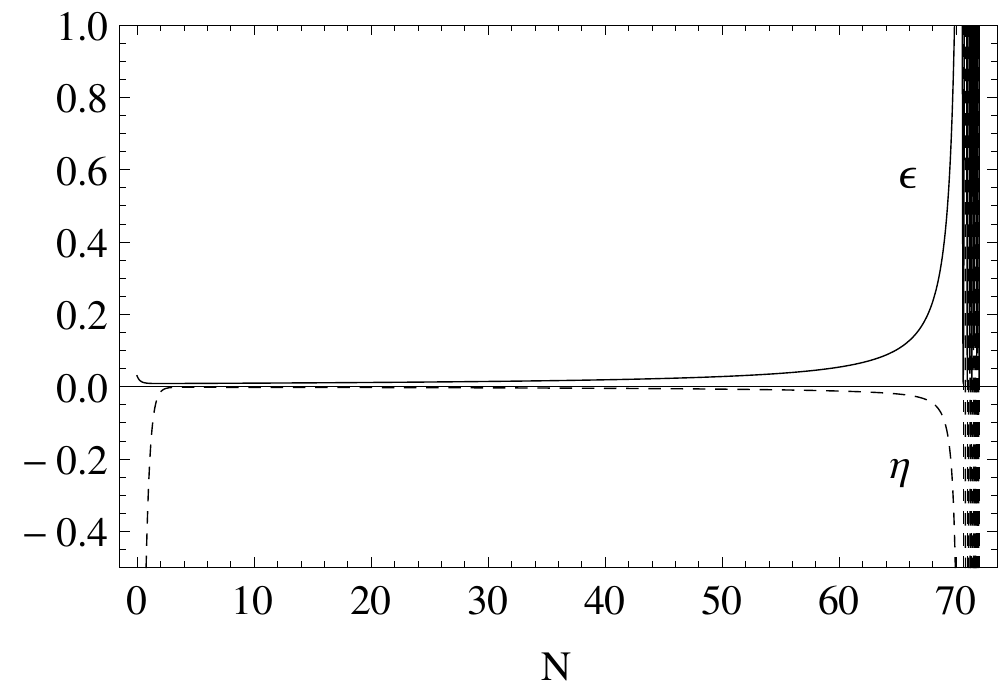}
\caption{In the left panel we have the graphical representation of the numerical
solutions of (\ref{2Quad-X-diff}) and (\ref{2Quad-Y-diff}) for $x_{1}\left(N\right)$
(full line) and $x_{2}\left(N\right)$ (dashed line) with $\theta=\dfrac{\pi}{4}$
for the potentials $V_{1}=x_{1}^{2}$ and $V_{2}=2x_{2}^{2}$. We
have taken the initial conditions as $x_{1}(0)=1.8\times\sqrt{\frac{1}{3}}$
and $x_{2}(0)=2.0\times\sqrt{\frac{1}{3}}$. In the right panel, and
for the same initial conditions, we have the graphical representation
of the numerical solutions for $\epsilon\left(N\right)$ (full line)
and $\eta\left(N\right)$ (dashed line).}

\label{fig7} 
\end{figure}

\begin{figure}[t]
\centering\includegraphics[height=2.5in]{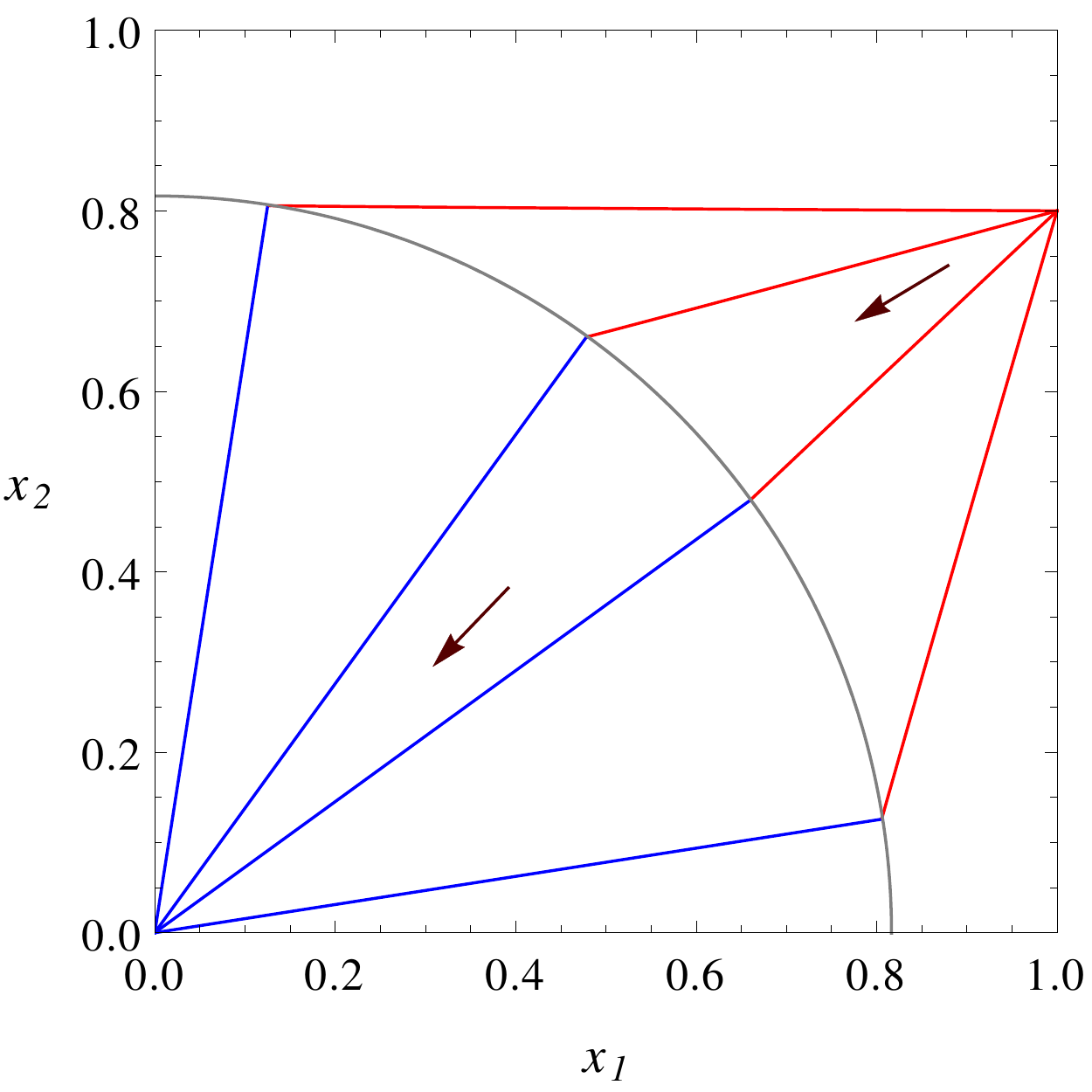}\quad{}\includegraphics[height=2.5in]{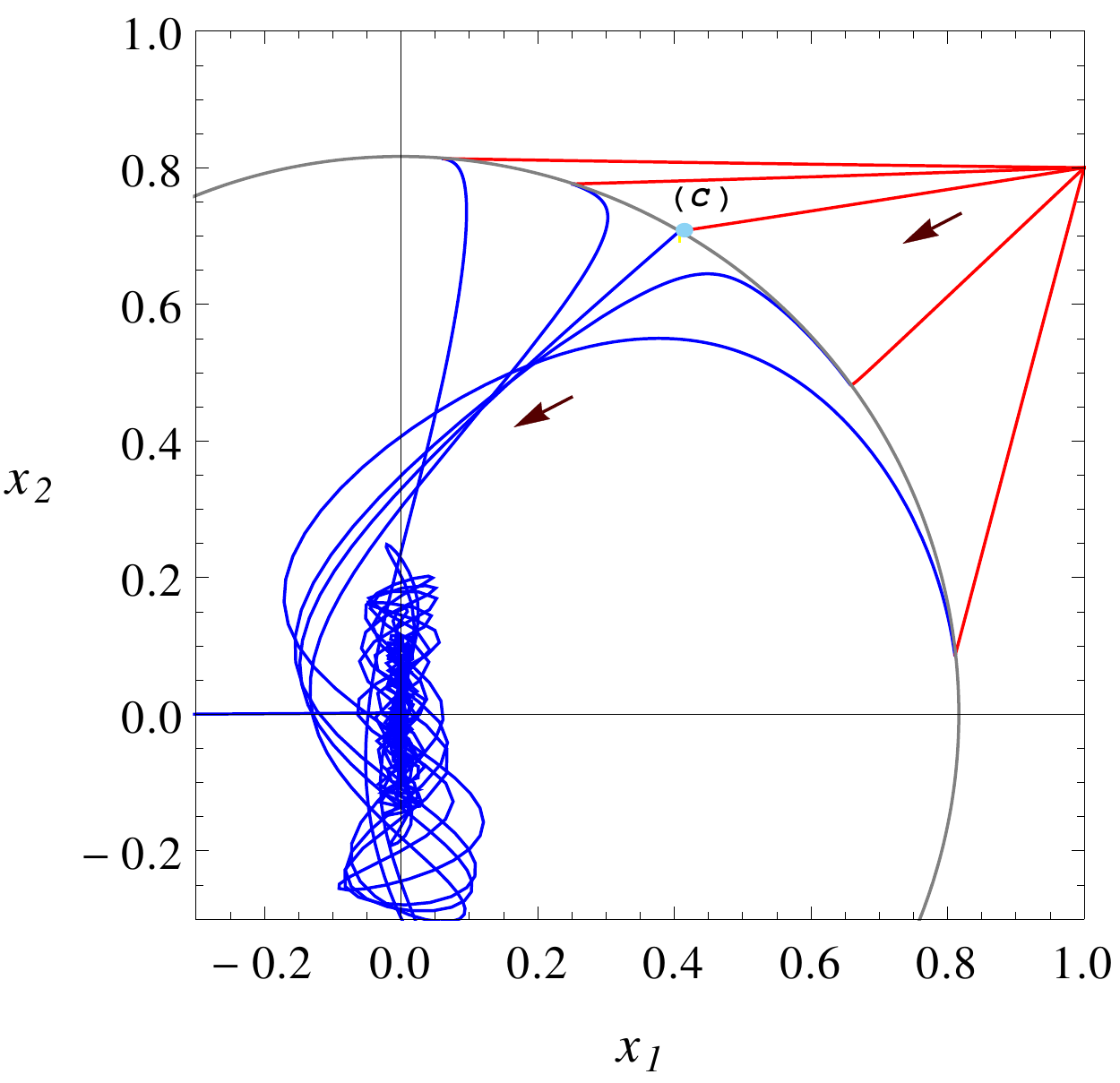}
\caption{This figure represents a set of trajectories evolving in the $\left(x_{1},\, x_{2}\right)$
space. These trajectories are numerical solutions of (\ref{2Quad-X-diff})
and (\ref{2Quad-Y-diff}) and correspond to a situation where we choose
$V_{1}=x_{1}^{2}$ and $V_{2}=x_{2}^{2}$ (left panel), as an illustrative
example only showing type I solution. All the fixed points are part
of the arc of radius $\sqrt{2/3}$ in the $\left(x_{1},x_{2}\right)$
plane. In the right panel, we have an example, where we have taken
$V_{1}=x_{1}^{2}$ and $V_{2}=x_{2}^{4}$, showing type II solutions,
except for the trajectory going close to a fixed point with $\theta=\pi/3$
(point $C$). In addition, in the right panel, we have an illustration
of two 3-form fields damped oscillations by the end of inflation.
The arrows, in the plots, indicate the direction of time in the trajectories. }

\label{fig7-sq-sq} 
\end{figure}

Let us now present the other class of inflationary
solution, which was mentioned in the Introduction. This type is associated
to the manner asymmetry is present. Let us be more specific. One way
to attain this solution consists of choosing an initial value of $\theta$
away from the fixed points previously discussed. This corresponds
to the curved trajectories in the right panel of Fig.\ref{fig7-sq-sq}.
Another is by choosing different sclales of the potentials. In any
case, the inflationary behavior (type II) is similarly affected concerning
either way of introducing asymmetry. We should note here that there
is no analog for a type II solution within single 3-form driven inflation.
To understand this new type of inflationary scenario, let us take
$V_{1}=x_{1}^{2}$ and $V_{2}=2x_{2}^{2}$ (just different slopes),
whose numerical solutions are plotted in Fig. \ref{fig7}.

In Fig. \ref{fig7}, the two fields continuously evolve, and at the
same time assist each other in order to sustain a slow roll regime.
As we can see from the left panel of Fig. \ref{fig7}, one field continues
to slowly decrease (dashed line) and the other (full line) starts
to increase until it enters in an oscillatory regime. However, in
the right panel of Fig. \ref{fig7}, we see that the slow roll parameters
evolve (before oscillating) near to zero during the period of inflation.
Moreover, from Eq. (\ref{N-3F-eta-1}), the behavior of the two fields
are such that even with $x_{n}^{\prime}\not\approx0$, the slow roll
conditions are consistent with inflation. The fact is that the slow
roll parameter $\eta\rightarrow0$ is now due to the constraint (\ref{NonIdent-sl-cond}).
As previously mentioned, a rather unusual cooperation between the
two 3-form fields, emphasized by the mentioned coupling (gravity mediated,
through $\dot{H}$) provides a different inflationary dynamics.

This new type of solution presents a period of inflation with an interesting
new feature. More precisely, when one 3-form field decreases, say
$x_{1}$, then the other field, $x_{2}$, is constrained to increase.
However, the increase of the second 3-form field is limited by the
fact that, as the first one inevitably approaches zero, then Eq. (\ref{Dyn-w2})
becomes 
\begin{equation}
w'_{2}\sim\frac{3}{2}\left(1-w_{2}^{2}\right)\lambda_{2}\left(x_{2}w_{2}-\sqrt{\frac{2}{3}}\right),\label{Dyn-w2-app1}
\end{equation}
with the coupling term $\lambda_{1}x_{1}w_{2}$ being negligibly small.
We see that Eq. (\ref{Dyn-w2-app1}) will become zero when $w_{2}$
(which is increasing, as is $x_{2}$) will approach $1$. At this
stage, and inspecting Eq. (\ref{Dyn-x2}), it is clear that $x_{2}$
will stop increasing and start to decrease, making $x_{2}'<0$. This
situation is depicted in the left panel Fig. \ref{fig7}, where the
decreasing field is reaching zero at the same period where the other
stops to increase and also converges to zero. The two 3-form fields
behave strongly correlated and assisting each other through the inflationary
period. Therefore, this more complex and correlated evolution of the
fields can provide a different observational signature when compared
to other multifield inflationary models.

The different nature of type I and type II solutions is be represented
in Fig.~\ref{fig7-sq-sq}. Therein, we have a parametric plot%
\footnote{Please note that Fig.~\ref{fig7-sq-sq} is $\mathit{not}$ a phase
space representation.%
} of $x_{1}(N)$ and $x_{2}(N)$ in the field space, where the fixed
points (cf. in particular the analysis in \ref{Quadrid} and \ref{Quadr-quartic})
are located at a pair of coordinates $(x_{1c},\, x_{2c})$, of course
associated to a situation where $(x_{1}',\, x_{2}')=0$. 

The two fields rapidly evolve towards this pair of coordinates, (cf.
the behavior illustrated in Figs. \ref{fig1} and \ref{fig7}) settling
there for the inflationary period. Afterwards, and because these fixed
points are not stable, the two fields will eventually diverge from
it. More precisely, in the left panel of Fig.~\ref{fig7-sq-sq} we
have the particular case where the two 3-form fields are under the
influence of identical quadratic potentials. In this case, only type
I solutions are present and the inflationary epochs, occur near the
depicted circle. Those fixed points in this figure are all located
in the arc of radius $\sqrt{2/3}$ in the $\left(x_{1},x_{2}\right)$
plane. The right panel, of the same figure, constitutes an example
where only one fixed point is present (using Eq. (\ref{cond-V1-V2}))
between $\theta=0$ and $\theta=\pi/2$. This fixed point, located
at $(C)$ in the right panel, corresponds to a type I solution when
$\theta=\pi/3$, for a case where the potentials are $V\left(x_{1}\right)=x_{1}^{2}$
and $V\left(x_{2}\right)=x_{2}^{4}$. All the other depicted trajectories
are type II solutions, where the $\dot{H}$-term coupling mediation
plays a crucial role(cf. Fig \ref{fig7}). The peculiar oscillatory
regime, present the right panel of \ref{fig7-sq-sq}, is also characteristic
of the coupling term in the effective potential (\ref{NGen-X-Pot-1}).
We shall discuss the oscillatory behavior in the following subsection.

\subsubsection{Oscillatory regime after inflation}

\textcolor{blue}{\label{oscillation} }The main purpose of this section
is to present an analytical description of the oscillatory behavior,
emerging by the end of inflation for the choice of potentials presented
in Table \ref{potential-stability}. This analysis can also be useful
for subsequent studies on reheating and particle production, as modeled
by the two 3-forms scenario which we postpone for a future work. The
interesting aspect that happens with two 3-forms is due to the presence
of the $\dot{H}$ coupling term in the effective potential (\ref{NDiff-syst-1-1-2}),
which becomes particularly dominant and produces a nontrivial interaction
between the 3-form fields in the type II case. At this point, we must
note that this property is more general, in the sense that the conclusion
drawn for two fields can be easily extended when more 3-form fields
are included. The choice of potential plays an important role regarding
the presence of a consistent oscillatory behavior, which successfully
avoid ghost instabilities by the end of inflation. This is illustrated
for single 3-form inflation in the Ref.~\cite{Felice-1,felice2}.
Based on the studies of single 3-form inflation, we chose potentials
containing quadratic behavior. Moreover, we must emphasize that the
oscillatory regime for two 3-forms case is different from single 3-form
inflation, due to the presence of the coupling term in the equations
of motion. An exception is the case of identical quadratic potentials,
i.e., taking $V_{(n)}=x_{n}^{2}$, where we can reasonably ignore
the effect of coupling. This is the special case where two 3-form
fields oscillate almost independently.

To illustrate this, let us first consider that the two fields are
subjected to quadratic potentials $V_{(n)}=\frac{1}{2}m_{n}^{2}\chi_{n}^{2}$.
For simplicity we work with the equations of motion in $t$ time (\ref{NDiff-syst-1-1}).
The equation of motion (\ref{NDiff-syst-1-1}) for the 3-form field
$\chi_{n}$ can be approximated in the small field limit $\left(\chi_{n}\rightarrow0\right)$
by neglecting the effect of coupling term in the effective potential
(\ref{NGen-X-Pot-1}) as, %
\begin{equation}
\ddot{\chi}_{n}+3H\dot{\chi}_{n}+m_{n}^{2}\chi_{n}\approx0.\label{quad-oscil-1}
\end{equation}
From the Friedmann constraint (\ref{NFriedm-1-1}) we have that during
inflation $H$ slowly decreases, since $\dot{H}<0$. When inflation
ends, $m_{n}^{2}\thicksim H^{2}$, and subsequently the 3-form fields
begin to coherently oscillate at scales $m_{n}^{2}\gg H^{2}$. The
evolution of $\chi_{n}$ at the oscillatory phase can be studied by
changing the variable $\chi_{n}=a^{-3/2}\bar{\chi}_{n}$, so that
Eq.~(\ref{quad-oscil-1}) becomes 
\begin{equation}
\ddot{\bar{\chi}}_{n}+\left(m_{n}^{2}-\frac{9}{4}H^{2}-\frac{3}{2}\dot{H}\right)\bar{\chi}_{n}\approx0\label{dampxnosci}
\end{equation}
Using the approximations $m_{n}^{2}\gg H^{2}$ and $m_{n}^{2}\gg\dot{H}$,
the solution to the Eq.~(\ref{dampxnosci}) can be written as 
\begin{equation}
\bar{\chi}_{n}=C\sin\left(m_{n}t\right)\label{xnbarsol}
\end{equation}
where $C$ is a the maximum amplitude of the oscillations. Thus the
solution for $\chi_{n}$ can be written as 
\begin{equation}
\chi_{n}=Ca^{-3/2}\sin\left(m_{n}t\right).\label{dampxnsolution}
\end{equation}

An interesting aspect arises in the small field limit when one of
the two 3-form fields potentials is not quadratic. Let us suppose
the situation described in \ref{Quadr-quartic}, with one field subjected
to a quartic potential, $V_{2}=\lambda\chi_{2}^{4}$. This discussion
is related to the oscillatory phase we see in the right panel of Fig.
\ref{fig7-sq-sq}, regarding the type II case. This combination of
potentials has the peculiar feature to induce an oscillatory regime,
more precisely, that for a single 3-form field it would be absent
under the quartic potential due to the presence of a ghost term \cite{Felice-1}.
In the limit $\chi_{1},\,\chi_{2}\rightarrow0$, towards the oscillatory
phase, the field $\chi_{1}$ will be approximately described by Eq.
(\ref{dampxnsolution}). Therefore the 3-form field $\chi_{1}$ undergoes
a damped oscillatory regime due to the dominance of quadratic behavior.
However, the second field $\chi_{2}$, also undergoes an oscillatory
regime, not caused by the quartic potential but due to the coupling
term, $V_{2,x_{2}}^{\textrm{eff}}$, dominance in Eq. (\ref{2Quad-Y-diff}).
The equation of motion (\ref{2Quad-X-diff}) for the 3-form field
becomes (in the small field limit, $\chi_{1},\,\chi_{2}\rightarrow0$,
near the oscillatory phase), 
\begin{equation}
\ddot{\chi}_{2}+3H\dot{\chi}_{2}+\left(4\lambda\chi_{2}^{3}-\dfrac{3}{2}m_{1}^{2}\chi_{1}^{2}\chi_{2}\right)\approx0.\label{quad-oscil-2}
\end{equation}
The nonlinear differential equation (\ref{quad-oscil-2}) is explicitly
affected by the oscillatory behavior of $\chi_{1}$, which could cause
something similar to a parametric resonance effect in particle production
\cite{felice2}. The effective potential also carries a cubic term,
which turns the equation difficult to solve. However, we can conjecture
that for two 3-forms inflation, at least one of the potentials must
contain a quadratic behavior, which forces all the other fields to
undergo a consistent oscillatory phase due to the influence of the
coupling term. In the case of the single 3-form inflation, there is
no oscillatory behavior for quartic potential, a fact that the authors
in \cite{Felice-1} explain by means of ghost instabilities. Therefore,
to conclude this section we present a new choice of potential i.e.,
$V_{1}=x_{1}^{2}$ and $V_{2}=x_{2}^{4}$, which can avoid ghost instabilities
due to the presence of consistent oscillatory phase. A similar oscillatory
regime is present when assisted inflation with two scalar fields is
studied by means of an explicit quartic coupling in the action \cite{Braden:2010wd}.

\subsubsection{Varying speed of sound for two 3-form fields}

\label{sec: sound speed} In the following we examine how the type
II solutions establish pressure perturbations with varying speed of
sound.

Adiabatic perturbations are defined by

\[
\frac{\delta p}{\dot{p}}=\frac{\delta\rho}{\dot{\rho}}
\]
where $p$ and $\rho$ are the pressure and energy density of the
system. Pressure perturbations can in general be expanded as a sum
of an adiabatic and a non adiabatic perturbations ($\delta p_{nad}$),
which is given by \cite{christ-houston}

\[
\delta p=\delta p_{{\rm nad}}+c_{s}^{2}\delta\rho
\]
where $c_{s}^{2}$= $\dot{p}/\dot{\rho}$ is the adiabatic sound speed
for scalar perturbations in a thermodynamic system. The distinction
between adiabatic sound speed and phase sound is given for scalar
field models in Ref. \cite{kmalikspeeds}\emph{. }When an adiabatic
system is composed with multiple scalar fields $\phi_{n}$, we have
that

\begin{equation}
\frac{\delta\phi_{i}}{\dot{\phi_{i}}}=\frac{\delta\phi_{j}}{\dot{\phi_{j}}}.\label{multadia}
\end{equation}
The condition (\ref{multadia}) is consequently valid for any two
scalar field systems. The above condition can also be applicable for
a system of $\mathbb{N}$ 3-forms because its action can (at least
formally) always be dualized and reduced to an action with $\mathbb{N}$
non canonical scalar fields \cite{NunesNG}.

The general expression for the adiabatic sound speed for $\mathbb{N}$
3-form fields is defined as

\begin{equation}
c_{s}^{2}=\frac{\dot{p}_{\mathbb{N}}}{\dot{\rho}_{\mathbb{N}}}.\label{speeddef}
\end{equation}
If we take (\ref{Ndens-1-1}) and (\ref{Npress-1-1}) within the slow
roll approximation $\chi''_{n}\ll V_{n}(\chi_{n}),$ we get, generally

\begin{equation}
c_{s}^{2}=\frac{\sum_{n=1}^{\mathbb{N}}\chi'_{n}\,\chi_{n}\, V_{,\chi_{n}\chi_{n}}}{\sum_{n=1}^{\mathbb{N}}\chi'_{n}\, V_{,\chi_{n}}}\label{speedN3form}
\end{equation}
Which, in the two 3-forms case, allows the speed of sound to be explicitly
written as

\[
c_{s}^{2}=\frac{\chi'_{1}\,\chi_{1}\, V_{,\chi_{1}\chi_{1}}+\chi'_{2}\,\chi_{2}\, V_{,\chi_{2}\chi_{2}}}{\chi'_{1}\, V_{,\chi_{1}}+\chi'_{2}\, V_{,\chi_{2}}}
\]
Unlike the one 3-form sound speed, in a two 3-forms setting the sound
speed will depend on $\chi'_{n}$. For type I inflation, for which
we have $\left(\chi'_{n}\approx0\right)$, the speed of sound (\ref{speedN3form})
becomes constant during inflation. For the type II solution, where
we have $\chi'_{n}\not\approx0$, the speed of sound, $c_{s}^{2}$,
can vary during the inflationary period. This varying speed can subsequently
exhibit a peculiar imprint in the primordial power spectrum, scale
invariance and bi-spectrum extracted from the CMB data. We are going
to explore, in the next two sections, observational consequences,
due to a varying speed of sound, upon important quantities like the
tensor-scalar ratio, spectral index and running spectral index, by
examining particular type II solutions for suitable choice of potentials.

\section{Isocurvature perturbations and primordial spectra}

\label{powerspectra} One important feature of multiple field models
is the generation of isocurvature perturbations. In this section we
examine the effect of these perturbations in the context of two 3-form
fields scenario. More concretely, we will distinguish, type I and
type II solutions, with respect to the evolution of isocurvature perturbations.

As depicted, in the right panel of Fig. \ref{fig7-sq-sq} type I solutions
are characterized by a straight line, whereas type II solutions follow
a curved trajectory in field space. In scalar multifield models, a
local rotation in the field space is carried to define the adiabatic
and entropy modes (or fields \cite{Gordon:2000hv}). In order to express
these adiabatic and entropy fields from two 3-form fields, the first
step consists in defining a dual scalar field Lagrangian for the two
3-forms (see Ref. \cite{NunesNG}). The second step consists in applying
the general framework of adiabatic and entropy perturbations to our
model. We start by presenting the $\mathbb{N}$ 3-forms Lagrangian
in terms of a non canonical Lagrangian $P\left(X,\phi_{n}\right)$,
where%
\footnote{The notation ``$n$'' indicating $n^{th}$field here corresponds
to ``$I$'' in Ref.\cite{Langlois:2008mn} %
} $X=-\frac{1}{2}\partial^{\mu}\phi_{n}\partial_{\mu}\phi_{n}$%
\footnote{Note that in the Ref.\cite{NunesNG} the definition $X=-\partial^{\mu}\phi_{n}\partial_{\mu}\phi_{n}$
differs by a $2$ factor from our notation. Accordingly, we have considered
this difference throughout this section. %
}, and $\phi_{n}$ is the dual scalar field corresponding to the 3-form
field tensor $F_{\alpha\beta\gamma\delta}$. The motivation to work
with the dual action is related to the fact that the general framework
of adiabatic and entropy perturbations for the non canonical multifield
model has already been consistently established. In the following
we will briefly review and adopt to our case the results described
previously in \cite{Langlois:2008mn,Arroja:2008yy,NunesNG,LangloisDBI:2009ej,Kaiser:2012ak}.

The action for the dual scalar fields representation of the $\mathbb{N}$
3-forms can be written as \cite{NunesNG}, 
\begin{equation}
P\left(X,\phi_{n}\right)=\overset{\mathbb{N}}{\underset{n=1}{\sum}}\left(\chi_{n}V_{n,\chi_{n}}-V\left(\chi_{n}\right)-\frac{\phi_{n}^{2}}{2}\right),\label{dual action}
\end{equation}
with 
\begin{equation}
X=\underset{n}{\sum}X_{n},\hspace{1cm}\textrm{and}\hspace{1cm}X_{n}=\frac{1}{2}V_{n,\chi_{n}}^{2}.\label{kinetic term}
\end{equation}
In the above equations, $\chi_{n}$ represent one element of the $\mathbb{N}$
3-form fields (cf. eqs. (\ref{N3f-Maxw-1})-(\ref{NDiff-syst-1-1})).
In the Lagrangian (\ref{dual action}) we can identify the kinetic
term to be 
\begin{equation}
K_{n}(X_{n})=\overset{\mathbb{N}}{\underset{n=1}{\sum}}\left(\chi_{n}V_{n,\chi_{n}}-V(\chi_{n})\right).\label{kinetic term-1}
\end{equation}
Since this kinetic term is only a function of $X_{n}$ and not of
$\phi_{n}$, this means that the field metric is $G_{nm}(\phi)=\delta_{nm}$.
Restricting ourselves now to a two 3-form scenario, and according
to \cite{Gordon:2000hv}, we can define the adiabatic and entropy
fields through a rotation in the two 3-form dual field space, 
\begin{equation}
\dot{\sigma}=\sqrt{2X_{1}}\,\cos\Theta+\sqrt{2X_{2}}\,\sin\Theta\label{adiab-field1}
\end{equation}
\begin{equation}
\dot{s}=-\sqrt{2X_{1}}\,\sin\Theta+\sqrt{2X_{2}}\,\cos\Theta\label{entrop-field1}
\end{equation}
where $\tan\Theta=\sqrt{X_{2}}/\sqrt{X_{1}}$, $X_{1}=\frac{1}{2}V_{1,\chi_{1}}^{2}$
and $X_{2}=\frac{1}{2}V_{2,\chi_{2}}^{2}$. Subsequently, the adiabatic
and entropy perturbations are 
\begin{equation}
Q_{\sigma}=\delta\phi_{1}\,\cos\Theta+\delta\phi_{2}\,\sin\Theta\label{adiab-perturb1}
\end{equation}
\begin{equation}
Q_{s}=-\delta\phi_{1}\,\sin\Theta+\delta\phi_{2}\,\cos\Theta,\label{entrop-perturb1}
\end{equation}
respectively, along and orthogonal to the background classical trajectory
in dual field space.

Let us assume that the linearly perturbed metric is given by 
\[
ds^{2}=-(1+2\varphi)dt^{2}+2\partial_{i}\beta dx^{i}dt+a^{2}(t)\left(1-2\psi\right)dx^{2}
\]
We have chosen a flat gauge, where the dynamics of linear perturbations
are completely expressed in terms of the scalar field perturbations
$\left(\phi^{n}\rightarrow\phi_{0}^{n}+Q^{n}\right)$. Moreover, these
are defined as gauge invariant combinations given by $Q^{n}=\delta\phi^{n}+\left(\phi^{n}/H\right)\psi$.
The comoving curvature perturbation is given by 
\begin{equation}
\mathcal{R}\equiv\psi-\frac{H}{p+\rho}\delta q\,,
\end{equation}
where $\partial_{i}\delta q_{i}=\delta T_{i}^{0}$ and $\mathcal{R}$
purely characterizes the adiabatic part of the perturbations. The
variation of $\mathcal{R}$, in the flat gauge, is given by \cite{Langlois:2008mn}
\begin{equation}
\dot{\mathcal{R}}=\frac{H}{\dot{H}}\frac{c_{s}^{2}k^{2}}{a^{2}}\Psi+\frac{H}{\dot{\sigma}}\,\Xi\, Q_{s}\quad\textrm{with}\quad\Xi=\frac{1}{\dot{\sigma}P_{,X}}\left(\left(1+c_{s}^{2}\right)P_{,s}-c_{s}^{2}\dot{\sigma}^{2}P_{,Xs}\right),\label{rdot at all}
\end{equation}
where $\Psi$ is the Bardeen potential and 
\begin{equation}
P_{,s}=P_{,X}\dot{\sigma}\dot{\Theta},\hspace{1cm}\left(\begin{array}{c}
P_{,X\sigma}\\
P_{,Xs}
\end{array}\right)=\left(\begin{array}{cc}
\cos\Theta & \sin\Theta\\
-\sin\Theta & \cos\Theta
\end{array}\right)\left(\begin{array}{c}
P_{,X_{1}}\\
P_{,X_{2}}
\end{array}\right)\,.\label{Pxs3form}
\end{equation}
For a two 3-form dual Lagrangian, extracted from (\ref{dual action}),
we can express the above quantities as functions of the 3-form fields,
i.e., 
\begin{equation}
P_{,X}\equiv P_{,X_{1}}+P_{,X_{2}}=\frac{\chi_{1}}{V_{1,\chi_{1}}}+\frac{\chi_{2}}{V_{2,\chi_{2}}}\label{px3form}
\end{equation}
Using Eqs. (\ref{px3form}) and (\ref{Pxs3form}) we can simplify
$\Xi$, to obtain, 
\begin{equation}
\Xi=H\left(\left(1+c_{s}^{2}\right)\,\frac{d\Theta}{dN}-c_{s}^{2}\,\frac{\dot{\sigma}}{H}\,\frac{P_{,Xs}}{P_{,X}}\right).\label{cascadeNtime}
\end{equation}
The function $\Xi$ is a measure of the coupling between the entropy
and adiabatic modes.

\subsection{Type I inflation}

\label{type1 entropy.} In type I inflationary scenarios, where $\dot{\Theta}=0$
(as $\tan\Theta=\lambda_{2}/\lambda_{1}=x_{2}/x_{1}=$ constant in
the fixed point, cf. Eq.(\ref{fixed-P2}) and see Fig.~\ref{fig7-sq-sq}),
the classical trajectory is a straight line. This fact makes the first
term of $\Xi$, in Eq. (\ref{cascadeNtime}), to vanish.

On the other hand, the ratio $P_{,Xs}/P_{,X}$ can be expressed as
\begin{equation}
\frac{P_{,Xs}}{P_{,X}}=\frac{-\chi_{1}\,\sin\Theta+\chi_{2}\,\cos\Theta}{\chi_{1}V_{1,\chi_{1}}+\chi_{2}V_{2,\chi_{2}}}.\label{cascadezerofortype1}
\end{equation}
Expression (\ref{cascadezerofortype1}) vanishes for all type I solutions
since $\chi_{2}=\chi_{1}\left(\lambda_{2}/\lambda_{1}\right)=\chi_{1}\tan\Theta$.
In other words, there are no entropy perturbations sourcing the curvature
perturbations. We then recover the known relation for a single field
inflation 
\begin{equation}
\dot{\mathcal{R}}=\frac{H}{\dot{H}}\frac{c_{s}^{2}k^{2}}{a^{2}}\Psi,
\end{equation}
and we can state that the curvature perturbation is conserved on the
large scales. We can, therefore, compute the power spectrum of curvature
perturbations in terms of quantities values at horizon exit.

\subsection{Type II inflation}

For type II inflation, the aforementioned effects, namely of entropy
perturbations, can be present due to the curved trajectory (cf. the
right panel of Fig.\ref{fig7-sq-sq}) in field space $(\dot{\Theta}\neq0)$.
Due to this the curvature power spectrum could be sourced by entropy
perturbations on large scales.

In order to study quantum fluctuations of the system we must consider
the following canonically normalized fields defined by, 
\[
v_{\sigma}=\frac{a\sqrt{P_{,X}}}{c_{s}}Q_{\sigma},\hspace{1cm}v_{s}=a\sqrt{P_{,X}}Q_{s}\,,
\]
we can express the second order action for the adiabatic and entropy
modes as 
\begin{equation}
S_{(2)}=\frac{1}{2}\int d\tau d^{3}k\left[v_{\sigma}^{\prime^{2}}+v_{s}^{\prime^{2}}-2\xi v_{\sigma}^{\prime}v_{s}-k^{2}c_{s}^{2}v_{\sigma}^{2}-k^{2}v_{s}^{2}+\Omega_{\sigma\sigma}v_{s}^{2}+\Omega_{ss}v_{\sigma}^{2}+2\Omega_{s\sigma}v_{\sigma}v_{s}\right]\label{2ndorder action}
\end{equation}
with 
\[
\xi=\frac{a}{c_{s}}\Xi,\quad\Omega_{\sigma\sigma}=\frac{z^{\prime\prime}}{z}\quad\textrm{and}\quad\Omega_{ss}=\frac{\alpha^{\prime\prime}}{\alpha}-a^{2}\mu_{s}^{2},
\]
where $z$ and $\alpha$ are background dependent functions defined
by 
\[
z=\frac{a\dot{\sigma}\sqrt{P_{,X}}}{c_{s}H},\;\alpha=a\,\sqrt{P_{,X}}\,.
\]
The equations of motion derived from the action (\ref{2ndorder action})
are given by 
\begin{equation}
v_{\sigma}^{\prime\prime}-\xi v_{s}^{\prime}+\left(c_{s}^{2}k^{2}-\frac{z^{\prime\prime}}{z}\right)v_{\sigma}-\frac{\left(z\xi\right)^{\prime}}{z}v_{s}=0\:,\label{sigma eqn}
\end{equation}
\begin{equation}
v_{s}^{\prime\prime}+\xi v_{\sigma}^{\prime}+\left(k^{2}-\frac{\alpha^{\prime\prime}}{\alpha}+a^{2}\mu_{s}^{2}\right)v_{s}-\frac{z^{\prime}}{z}\xi v_{\sigma}=0\:,\label{s eqn}
\end{equation}
where $\mu_{s}^{2}$ is the effective mass for the entropy field given
by \cite{Langlois:2008mn} 
\begin{equation}
\mu_{s}^{2}=-\frac{P_{,ss}}{P_{,X}}-\frac{1}{2c_{s}^{2}\left(X_{1}+X_{2}\right)}\frac{P_{,s}^{2}}{P_{,X}^{2}}+2\frac{P_{,Xs}P_{,s}}{P_{,X}^{2}}
\end{equation}
and 
\begin{equation}
\left(\begin{array}{cc}
P_{,\sigma\sigma} & P_{,\sigma s}\\
P_{,s\sigma} & P_{,ss}
\end{array}\right)=\left(\begin{array}{cc}
\cos\Theta & \sin\Theta\\
-\sin\Theta & \cos\Theta
\end{array}\right)\left(\begin{array}{cc}
P_{,X_{1}X_{1}} & P_{,X_{1}X_{2}}\\
P_{,X_{2}X_{1}} & P_{,X_{2}X_{2}}
\end{array}\right)\left(\begin{array}{cc}
\cos\Theta & -\sin\Theta\\
\sin\Theta & \cos\Theta
\end{array}\right)
\end{equation}
The coupling between adiabatic and entropy modes is governed by the
parameter $\xi$. In the cases where this parameter can be assumed
to be small (see \cite{Langlois:2008mn,Arroja:2008yy}) at the typical
scale of sound horizon exit %
\footnote{In contrast to the inflationary models where a sharp turn in field
space occurs during inflation \cite{Lalak:2007vi,Peterson:2010np,Konieczka:2014zja}.%
} the adiabatic and entropy modes decouple and analytical solutions
for Eqs. (\ref{sigma eqn})-(\ref{s eqn}) can easily be found. In
the decoupled case the adiabatic and entropy modes evolve according
to the following equations, 
\begin{equation}
v_{\sigma}^{\prime\prime}-\left(c_{s}^{2}k^{2}-\frac{z^{\prime\prime}}{z}\right)v_{\sigma}=0,\label{free adiabatic}
\end{equation}
\begin{equation}
v_{s}^{\prime\prime}+\left(k^{2}-\frac{\alpha^{\prime\prime}}{\alpha}+a^{2}\mu_{s}^{2}\right)v_{s}=0.\label{freeentropy}
\end{equation}

In the slow roll limit, for a speed of sound that slowly varies while
the scales of interest cross out the sound horizon, we can assume
$z^{''}/z^{'}=1/\tau^{2}$. Using this, we get as a general approximate
solutions for the adiabatic and entropy modes with Bunch-Davies vacuum
initial conditions, 
\begin{equation}
v_{\sigma k}\simeq\frac{1}{\sqrt{2kc_{s}}}\exp\left(-ikc_{s}\tau\right)\left(1-\frac{i}{kc_{s}\tau}\right),\label{adiabatic solution}
\end{equation}
\begin{equation}
v_{sk}\simeq\frac{1}{\sqrt{2k}}\exp\left(-ik\tau\right)\left(1-\frac{i}{k\tau}\right),\label{entropy solution}
\end{equation}
where we assume $\frac{\mu_{s}^{2}}{H^{2}}\ll1$ is valid for our
case. This means entropy modes get amplified with respect to the adiabatic
modes at the sound horizon crossing 
\begin{equation}
Q_{\sigma_{*}}\simeq\frac{Q_{s_{*}}}{c_{s_{*}}}.\label{entropy amp}
\end{equation}
The curvature and isocurvature perturbations are respectively, 
\begin{equation}
\mathcal{R}=\frac{H}{\dot{\sigma}}Q_{\sigma},\hspace{1cm}\mathcal{S}=c_{s}\frac{H}{\dot{\sigma}}Q_{s}\label{adiapentp}
\end{equation}
The power spectrum of the curvature perturbation, evaluated at the
sound horizon crossing $\left(c_{s}k=aH\right)$, is given by 
\begin{equation}
\mathcal{P}_{\mathcal{R}_{*}}=\frac{k^{3}}{2\pi^{2}}\frac{\mid v_{\sigma k}\mid^{2}}{z^{2}}\simeq\frac{H^{4}}{8\pi^{2}XP_{,X}}=\frac{H^{2}}{8\pi^{2}\epsilon c_{s}}\Bigg|_{*}\:,\label{curvature power spectrum}
\end{equation}
which recovers with the single field power spectrum result at horizon
crossing \cite{NunesNG}. However, in contrast to the single field
inflation, the function $\xi$ is not negligible and typically varies
with time. This means that there will be a transfer between entropic
and adiabatic modes on large scales but the converse is not true.
From Eqs. (\ref{rdot at all}) and (\ref{adiapentp}), the evolution
of the curvature and entropy modes in the long wavelength limit can
be approximated as \cite{LangloisDBI:2009ej} 
\[
\dot{\mathcal{R}}\approx\alpha HS,\hspace{1cm}\dot{\mathcal{S}}\approx\beta HS,
\]
where the coefficients $\alpha$ and $\beta$ are taken to be, 
\begin{equation}
\alpha=\frac{\Xi}{c_{s}H},\label{alpha}
\end{equation}
\begin{equation}
\beta\simeq\frac{s}{2}-\frac{\eta}{2}-\frac{1}{3H^{2}}\left(\mu_{s}^{2}+\frac{\Xi^{2}}{c_{s}^{2}}\right),\label{beta}
\end{equation}
endowed with the definition of an additional slow roll parameter $s=\frac{\dot{c}_{s}}{Hc_{s}}$.
The evolution of curvature and isocurvature perturbations after horizon
crossing can be evaluated using transfer functions defined by 
\[
\left(\begin{array}{c}
\mathcal{R}\\
\mathcal{S}
\end{array}\right)=\left(\begin{array}{cc}
1 & \mathcal{T}_{\mathcal{R}\mathcal{S}}\\
0 & \mathcal{T}_{\mathcal{S}\mathcal{S}}
\end{array}\right)\left(\begin{array}{c}
\mathcal{R}\\
\mathcal{S}
\end{array}\right)_{*},
\]
where 
\begin{equation}
\mathcal{T}_{\mathcal{R}\mathcal{S}}\left(t_{*},t\right)=\int_{t_{*}}^{t}dt^{\prime}\alpha\left(t^{\prime}\right)H\left(t^{\prime}\right)T_{SS}\left(t_{*}\right),\label{trs}
\end{equation}
and 
\begin{equation}
\mathcal{T}_{\mathcal{S}\mathcal{S}}\left(t_{*},t\right)=\exp\left\{ \int_{t_{*}}^{t}dt^{\prime}\beta\left(t^{\prime}\right)H\left(t^{\prime}\right)dt^{\prime}\right\} ,\label{tss}
\end{equation}
In addition, the curvature perturbation power spectrum, the entropy
perturbation and the correlation between the two can be formally related
as 
\begin{equation}
\mathcal{P_{R}=}\left(1+\mathcal{T}_{\mathcal{R}\mathcal{S}}^{2}\right)\mathcal{P}_{*},\hspace{1cm}\mathcal{P_{S}=}\mathcal{T}_{\mathcal{S}\mathcal{S}}^{2}\mathcal{P}_{*},\label{full powerspectrum}
\end{equation}
\begin{equation}
\mathcal{C_{RS}}\equiv\langle\mathcal{RS}\rangle=\mathcal{T}_{\mathcal{R}S}\mathcal{T_{SS}}\mathcal{P}_{*}.\label{coupling power}
\end{equation}
In contrast to the power spectrum for the scalar perturbations, the
tensor power spectrum amplitude is the same as for a single field,
\[
\mathcal{P}_{\mathsf{T}}=\frac{2}{\pi^{2}}\frac{H^{2}}{M_{PI}^{2}}\Bigg|_{*}.
\]
The tensor scalar ratio defined in multifield inflation is given by

\begin{equation}
r\equiv\frac{\mathcal{P}_{\mathsf{T}}}{\mathcal{P_{R}}}=16\epsilon c_{s}\Bigg|_{*}\cos^{2}\Delta\:,\label{modified tensor scalar}
\end{equation}
where $\Delta$ is the transfer angle given by 
\[
\cos\Delta=\frac{1}{\sqrt{1+\mathcal{T}_{\mathcal{R}\mathcal{S}}^{2}}}.
\]
Similarly, the spectral index also gets a correction, provided by
the transfer functions, 
\begin{equation}
n_{s}\equiv\frac{d\,\ln\mathcal{P_{R}}}{d\,\ln k}=n_{s}(t_{*})+\frac{1}{H_{*}}\left(\frac{\partial T_{RS}}{\partial t_{*}}\right)\sin\left(2\Delta\right),\label{modifiedns}
\end{equation}
where 
\[
n_{s_{*}}=1-2\epsilon_{*}-\eta_{*}-s_{*}.
\]

The spectral index and the tensor to scalar ratio are the key observables
which not only depend on the slow roll at horizon crossing, but also
depend on the transfer angle $\Delta$. This enables a clear distinction
between multifields and single field inflationary scenarios %
\footnote{However tensor scalar ratio is more constrained by consistency relations
in case of inflation with more than two fields \cite{david-wand2,Bartolo:2001rt}.%
} \cite{david-wand2}. The transfer functions defined in Eqs. (\ref{trs})
and (\ref{tss}) are allowed to evolve after the Hubble exit, even
after inflation, during the reheating and radiation dominated era
\cite{david-wand2,Vernizzi:2006ve}. However the evolution of isocurvature
perturbations, during reheating and radiation dominated era, would
depend on the particular final stage of the inflationary scenario.
Consider for example, a two field scenario, if one field enters a
regime of oscillations while the second field is still inflating the
Universe. In such cases the curvature perturbation can be sourced
by entropy modes even after inflation \cite{Bartolo:2001rt}. This
kind of scenarios are known as `curvaton' or `spectator' field behavior
\cite{Elliston:2013afa,Choi:2008et} and also found in double quadratic
inflation \cite{Vernizzi:2006ve}. In the case of two 3-forms inflation,
we will assume that entropy perturbations do not grow further after
inflation. Therefore we only evaluate transfer functions from horizon
exit until the end of inflation and predict the values of $n_{s}$
and $r$ \cite{Peterson:2010np}. We can see from Eqs. (\ref{modified tensor scalar})
and (\ref{modifiedns}) that if $\mathcal{T}_{\mathcal{RS}}=0$ then
our predictions match the single field result. From the section \ref{type1 entropy.}
and Eq. \ref{alpha} it is evident that $\mathcal{T}_{\mathcal{RS}}=0$
for type I inflation. Therefore to make observational contrast with
single 3-form we mainly focus on testing type II inflationary scenario
in the following section.

\section{Two 3-form fields inflation and observational data}

\label{sec:Planck} Based on the discussion made on the curvature
perturbation power spectrum in section \ref{powerspectra}, the main
objective is to test our two 3-forms model and predicting values of
inflationary parameters. We choose suitable potentials and initial
conditions, in order to obtain a reasonable fit with present available
experimental bounds \cite{PconInflation,Ade:2014xna}. The majority
of inflationary models with a non canonical kinetic term contain a
common feature that the adiabatic fluctuations propagate with a sound
speed $c_{s}^{2}<1.$ The recent Planck data restricts this speed
of sound to be in the interval $0.02\lesssim c_{s}^{2}<1$. Multiple
field inflation models allow the possibility of having a varying speed
of sound, i.e, like for the type II solution in our model (cf. section
\ref{sec: sound speed}). The speed of sound variation will therefore
have implications on the running spectral index and the scale invariance.
These peculiar effects, being a consequence of the varying speed of
sound, have been studied in a DBI context and also in modified gravity
models with an effective inflaton \cite{khourypiazza,park-sorbo,fudenteass}.

We have examined all the potentials in Table \ref{potential-stability}.
We found that $\chi_{i}^{2}+b_{i}\chi_{i}^{4}$ is consistent with
observational bounds%
\footnote{We confront our results with $\chi_{i}^{2}+b_{i}\chi_{i}^{4}$ potential,
and one can find make similar predictions with $\chi_{i}^{2}+b_{i}\chi_{i}^{3}$
potential. We do not consider to explore quadratic potential as it
is equivalent to inflation with canonical scalar fields (in dual picture). %
}. It is quite difficult to constrain the speed of sound $\left(0.02\lesssim c_{s}^{2}<1\right)$
during inflation. We found that only type II solutions which are slightly
deviated from type I are suitable to maintain consistent speed of
sound during inflation. To predict values of inflationary parameters,
first we need to compute the transfer functions defined in section
\ref{powerspectra} and evaluate their value at the end of inflation.

\begin{figure}[ht!]
\centering\includegraphics[height=2.5in]{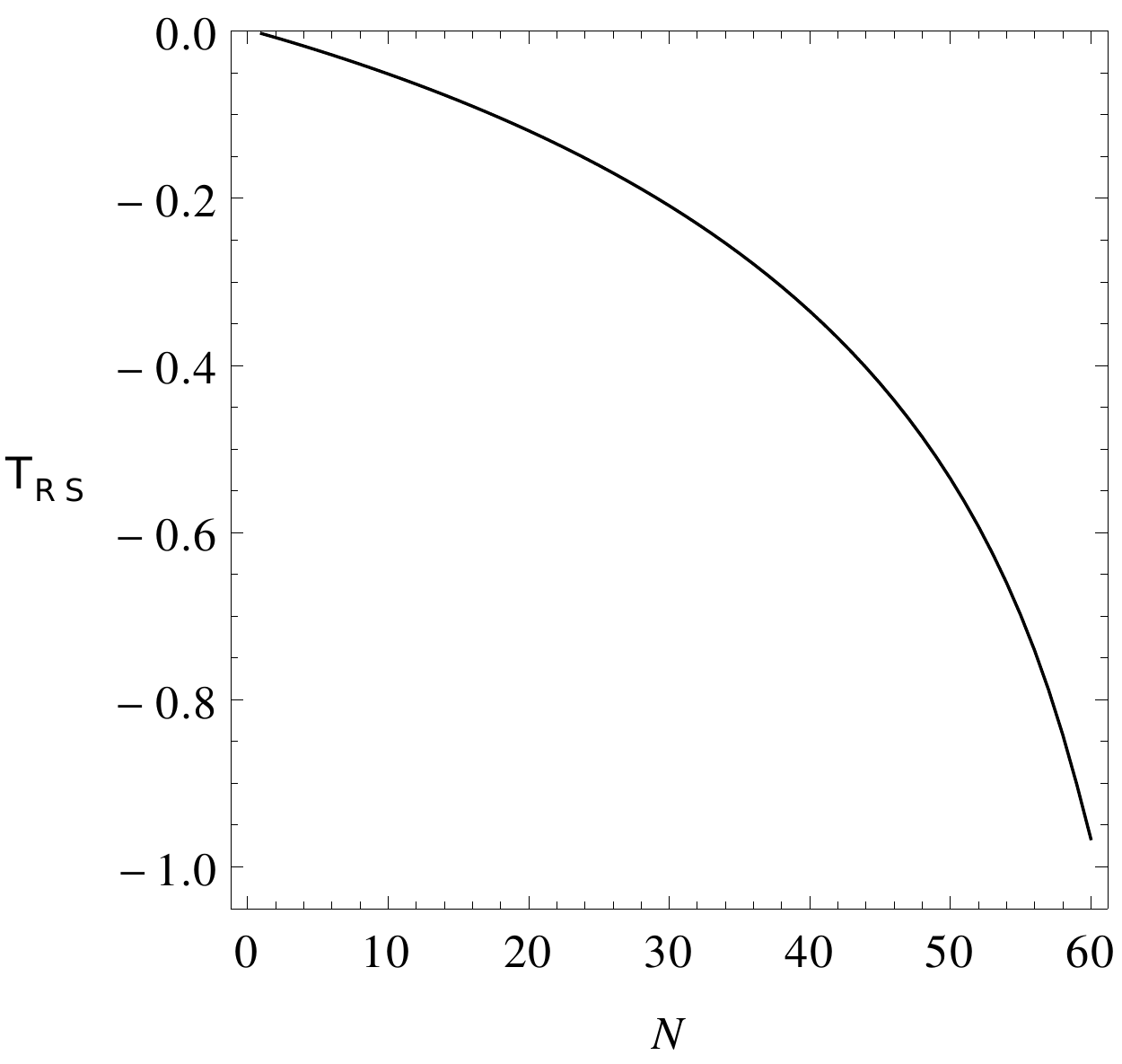}\quad{}\includegraphics[height=2.5in]{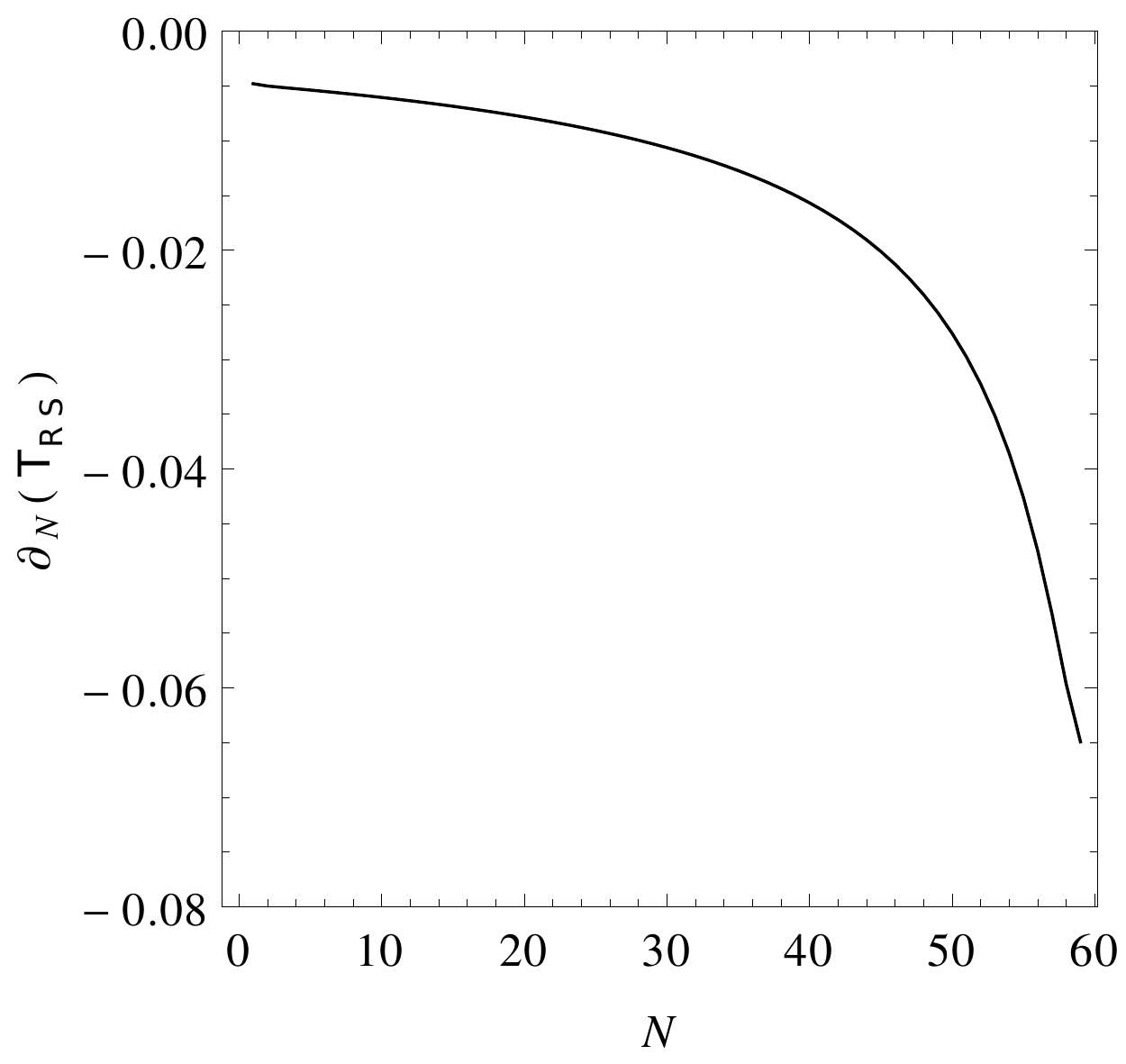}
\caption{Graphical representation of $\mathcal{T_{RS}}$ (left panel) and $\frac{d\mathcal{T_{RS}}}{dN}$
(right panel) till the end of inflation (defined for $\epsilon=1$).
We have taken $V_{1}=V_{10}(x_{1}^{2}+bx_{1}^{4})$ and $V_{2}=V_{20}(x_{2}^{2}+bx_{2}^{4})$
where $V_{10}=1,\: V_{20}=0.93,\: b=-0.35$ and with initial conditions
$\theta=\pi/4$.}

\label{tranferfunctions} 
\end{figure}

We can read from Eq. (\ref{modifiedns}) that the spectral index depends
on the derivative of $\mathcal{T_{RS}}$ at horizon crossing. From
the right panel of Fig. \ref{tranferfunctions} it is clear that the
derivative of $\mathcal{T_{RS}}$, between $N=0$ and $N=20$, is
very small and we can, therefore, neglect it. Hence, our prediction
of spectral index only depends on the values of the slow roll parameters
at horizon exit. 

The running of the spectral index to the lowest order in slow roll
is now given by, regime, 
\begin{equation}
\begin{aligned}\frac{dn_{s}}{d\ln k} & = & =\,\left(1+\epsilon+\frac{c_{s}^{\prime}}{c_{s}}\right)\Bigg|_{*}\left(n'_{s_{*}}+\frac{\partial\mathcal{T_{RS}}}{\partial N_{*}}\frac{\partial}{\partial N}\left(\frac{2\mathcal{T_{RS}}}{1+\mathcal{T}_{\mathcal{R}\mathcal{S}}^{2}}\right)+\frac{\partial^{2}\mathcal{T_{RS}}}{\partial N_{*}^{2}}\sin2\Delta\right).\end{aligned}
\label{Running sp}
\end{equation}
For the choice of potential in Fig.~\ref{tranferfunctions} we can
neglect the transfer function corrections to the running spectral
index (\ref{Running sp}). Therefore for this case the additional
slow roll parameter $s=\frac{c_{s}^{\prime}}{c_{s}}$ is of relevance,
which enables us to observationally distinguish between two 3-forms
and single 3-form inflation%
\footnote{In the single 3-form case \cite{Felice-1,Nunes-2} and also in the
type I solution of two 3-forms case, this additional slow roll parameter
satisfies, $s\equiv\frac{\dot{c_{s}}}{c_{s}H}=0.$%
}, with respect to the running of spectral index. Expression (\ref{Running sp})
is expanded up to the first order in the slow roll parameters. The
second order corrections are crucial if there is an abrupt path turn
in field space during horizon exit. These types of scenarios are considered
in detail in studies related with hybrid inflation and double quadratic
inflation \cite{Avgoustidis:2011em}. We can neglect these corrections
for two 3-form inflation, since the type II solutions herein considered
do not exhibit abrupt turns in field space under slow roll conditions.

To predict tensor scalar ratio (\ref{modified tensor scalar}) for
two 3-forms it is required to know the value of $\mathcal{T_{RS}}$
at the end of inflation. From the left panel of Fig. \ref{tranferfunctions},
$\mathcal{T_{RS}}$ is $\mathcal{O}(1)$ at the end of inflation.
Therefore it can reduce the value of tensor scalar ratio in contrast
to the single 3-form case.

Evidently two 3-forms inflation can be observationally distinguished
from single 3-form inflation, due to the possibility of a varying
speed of sound (cf. section \ref{sec: sound speed}) and transfer
function corrections by the end of inflation. Our method of observational
analysis are quite similar to the studies in \cite{Peterson:2010np,Kaiser:2012ak}.
In the following we confront our results against Planck+WP+BAO data
which provides $\frac{dn_{s}}{d\ln k}=-0.013\pm0.009$ for the running
of spectral index, and $\frac{d^{2}n_{s}}{d\ln k^{2}}=0.017\pm0.009$
for the running of running spectral index, both at 95\% CL, which
rules out exact scale-invariance at more than 5$\sigma$ level. Our
analysis show that for type II solution, a better fit can be achieved
given the current observational bounds (ruling out exact scale-invariance).

\begin{figure}[H]
\centering\includegraphics[height=2.15in]{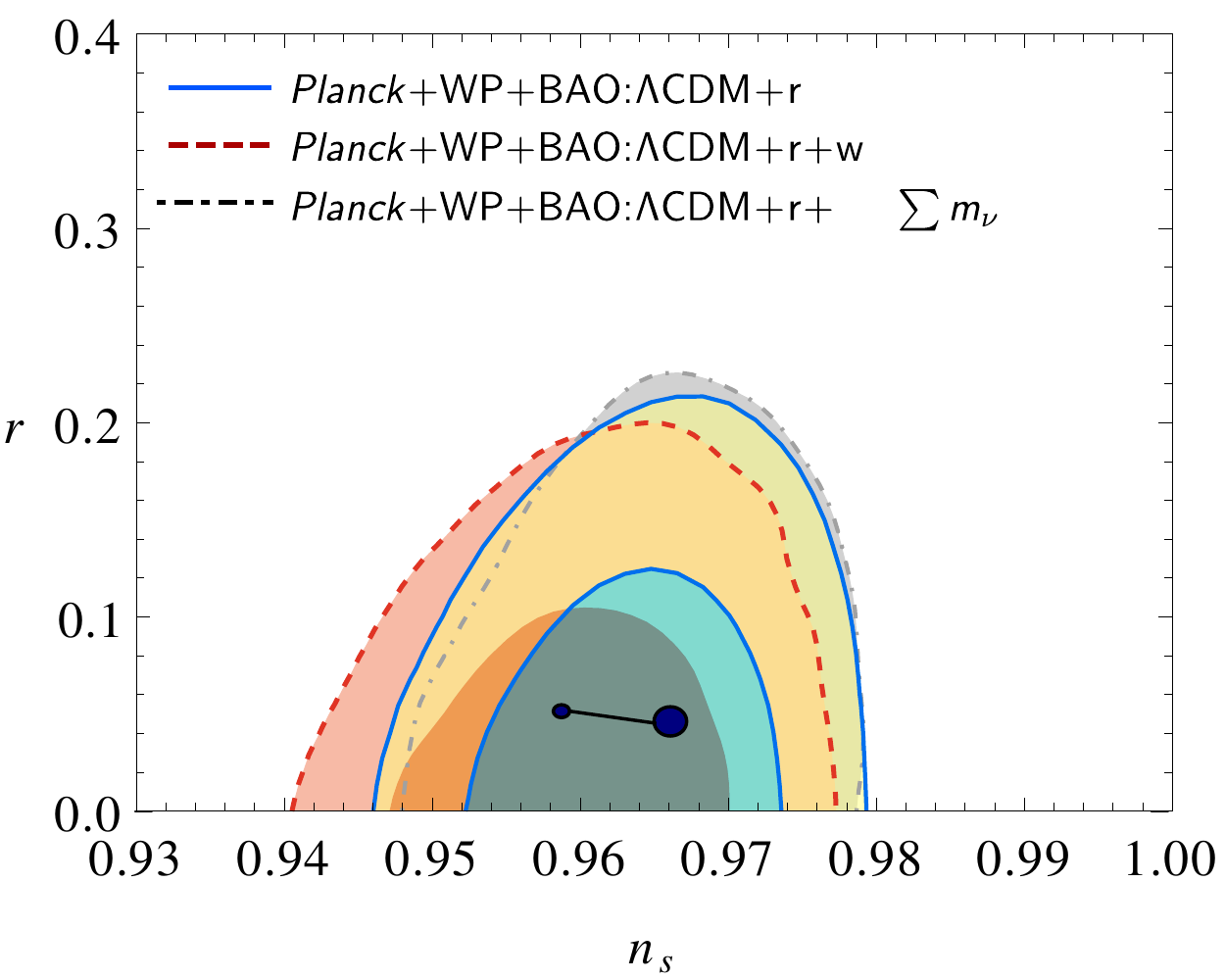}\quad{}\includegraphics[height=2.1in]{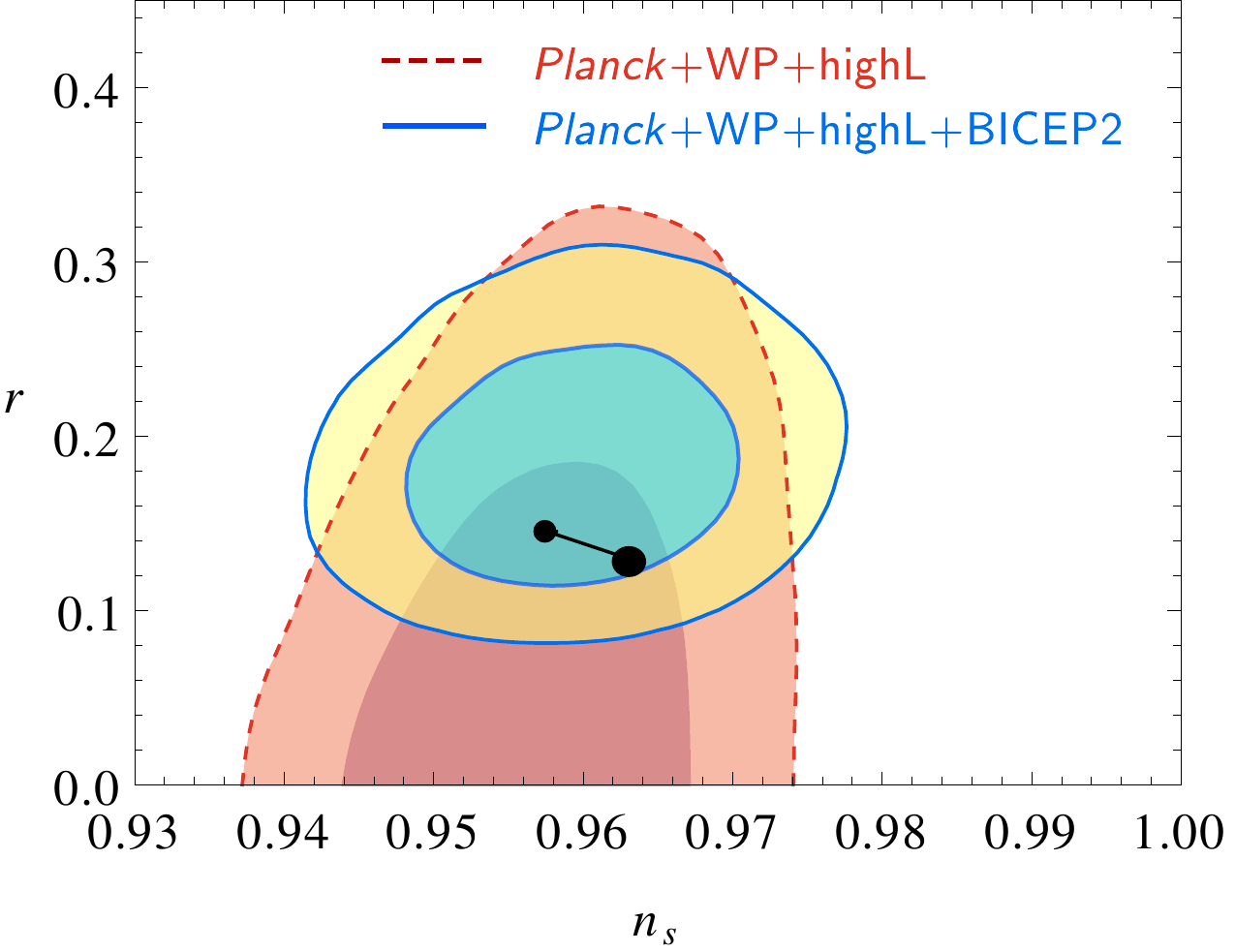}
\caption{Graphical representation of the spectral index versus the tensor to
scalar ratio, in the background of Planck+WP+BAO data (left panel),
for $N_{*}=60$ number of $e$-folds before the end of inflation (large
dot) and $N_{*}=50$ (small dot). We have taken $V_{1}=V_{10}(x_{1}^{2}+bx_{1}^{4})$
and $V_{2}=V_{20}(x_{2}^{2}+bx_{2}^{4})$ where $V_{10}=1,\: V_{20}=0.93,\: b=-0.35$
for two 3-form. The right panel corresponds to $n_{s}$ versus $r$
in the background of Planck+WP+highL+BICEP2 data. We have taken the
potentials $V_{1}=V_{10}(x_{1}^{2}+b_{1}x_{1}^{4})$ and $V_{2}=V_{20}(x_{2}^{2}+b_{2}x_{2}^{4})$
where $V_{10}=1,\: V_{20}=0.99,\: b_{1}=-0.14,\, b_{2}=-0.122.$ This
figure was obtained by taking the initial condition $\theta=\pi/4$.}

\label{fig8-1} 
\end{figure}

\begin{figure}[h]
\centering\includegraphics[height=2in]{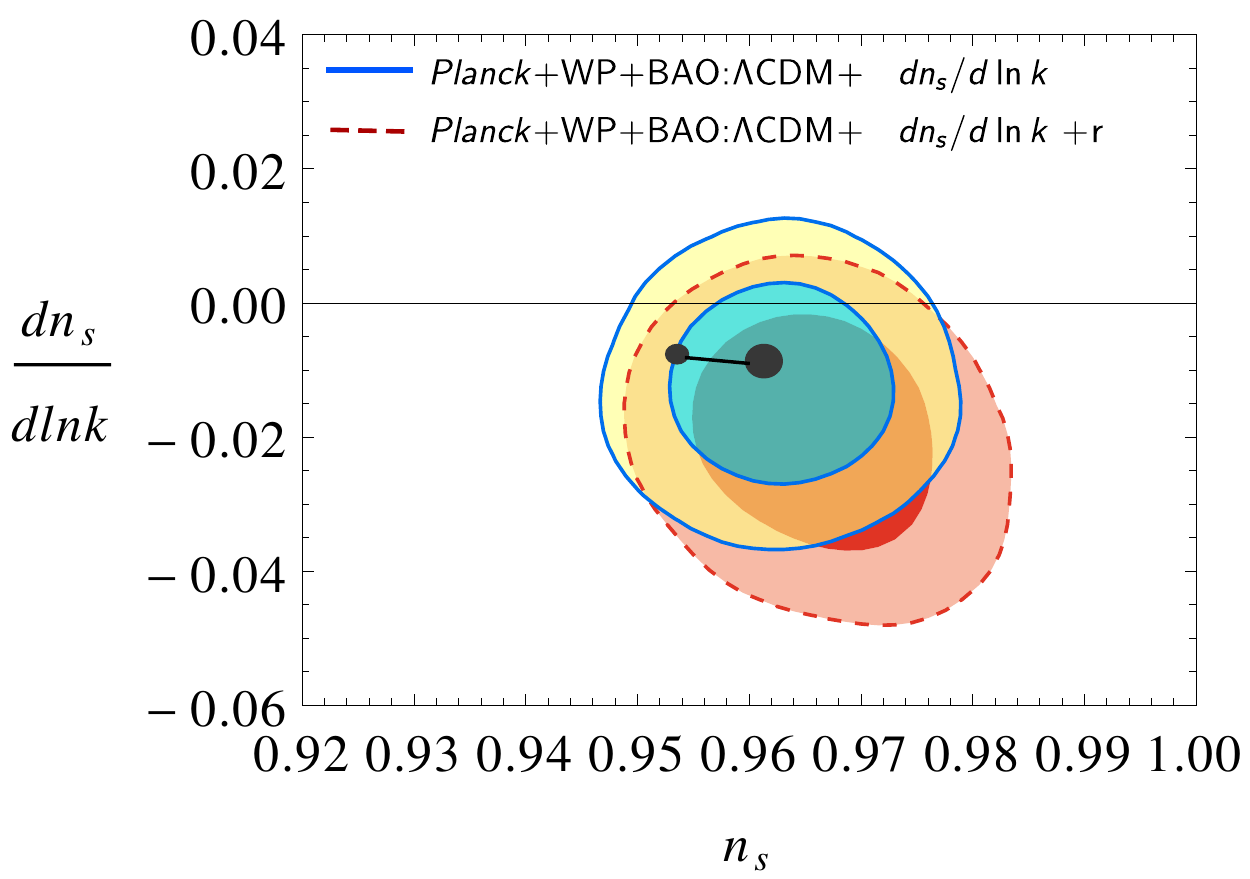}\quad{}\includegraphics[height=2in]{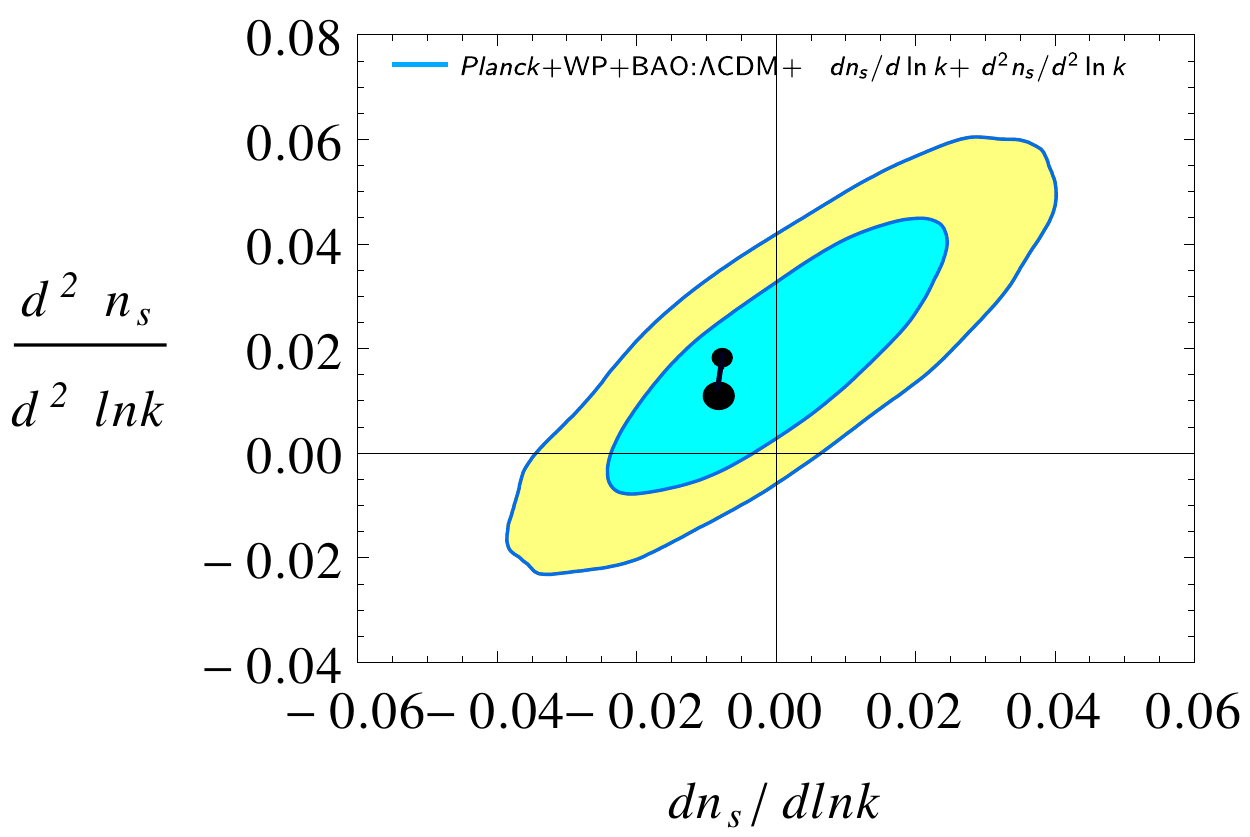}
\caption{Graphical representation of the running of the spectral index versus
the spectral index (left panel), and running of the running of the
spectral index versus the running of the spectral index (right panel)
in the background of Planck+WP+BAO data for $N_{*}=60$ number of
$e$-folds before the end of inflation (large dot) and $N_{*}=50$
(small dot). We have taken $V_{1}=V_{10}(x_{1}^{2}+bx_{1}^{4})$ and
$V_{2}=V_{20}(x_{2}^{2}+bx_{2}^{4})$ where $V_{10}=1,\: V_{20}=0.93,\: b=-0.35$
for two 3-form. This figure was also obtained by taking the initial
condition $\theta=\pi/4$.}

\label{fig8-2} 
\end{figure}

In Figs. \ref{fig8-1} and \ref{fig8-2}, obtained through suitable
data manipulating programs \cite{Lewis:2002ah,Lewis:2013hha}, we
have examined various types of potentials for a reasonable fit to
the observational constraints by Planck. In addition, we have also
considered the possibility of a higher tensor scalar ratio, since
the latest B-mode polarization analysis (BICEP2) discards a tensor
scalar ratio value $r=0$ at $7\sigma$ \cite{Ade:2014xna}. We found
that potentials such as $V_{i}=V_{i0}\left(\chi_{i}^{2}+b_{i}\chi_{i}^{4}\right)$
allow favorable contrast of two 3-forms inflation scenario against
recent observational data. The parameter $b_{i}$, in the mentioned
potential, is adequately chosen, so that the speed of sound gets bounded
by $0.02\lesssim c_{s}^{2}<1$, in order to comply with the Planck
constraint. We found that type II inflation, obtained through a small
asymmetry in the slopes of the potentials (making $V_{10}\neq V_{20}$),
is needed to fit the parameters within the bounds of the observational
data, especially for the running and running of running spectral indexes.
There are two relevant aspects that should be mentioned regarding
this comparison; one is related to the property of type II solution
for computing the running of the spectral index. This is a consequence
of the varying speed of sound, which is natural for this solution.
The other aspect is the requirement of the asymmetry between the potentials.
This leads to a mild generation of isocurvature perturbations towards
the end of inflation, which can accommodate tensor scalar ratio values
within the present bounds of Planck or with BICEP2 data as well. We
note that solutions with large curved trajectory in field space can
lead to values for inflationary parameters beyond the observational
bounds. The presence of curvature, in the field space trajectories,
implies a peculiar imprint in the primordial bispectrum during multiple
field inflation \cite{Kaiser:2012ak}. Therefore, a natural extension
of this work is to study the non-Gaussianity for two 3-form inflation
\cite{SJNP-NG}. 

\section{Conclusions, discussion and outlook}

\label{discussion} Let us summarize our specific results. Inflation
driven by a multifield setting, in particular by a couple of 3-form
fields, is very much still admissible within current Planck or BICEP2
data. This is the main assertion that this work indicates. Moreover,
two 3-form fields with a small asymmetry (in the sense explained in
this paper) produces better results (in terms of fitting within current
observational data) for concrete cosmological parameters, in contrast
to a symmetric configuration or to a single 3-form setting. This is
interesting if we take into consideration, the correspondence (on
dualization) between 3-form field and non-canonical (kinetical) scalar
field dynamics. In fact, a dual description of two 3-forms assists
to relate to k-inflationary models \cite{Ohashi:2011na}. First, we
have shown that having multiple 3-forms driven inflation brings the
inflaton mass to a lower scale, when compared with a single 3-form.
We then identified the existence of de Sitter like fixed points, where
two 3-forms inflation can mimic single 3-form inflationary scenarios,
for a suitable class of potentials. We also did a detailed numerical
study of a different type of inflationary dynamics (type II) characterized
by the dominance of a non trivial (gravity mediated) coupling, between
the two 3-form fields. The type II solution stands physically interesting
by its ability to generate substantial isocurvature perturbations
at the end of inflation. We have numerically computed the effect of
these perturbations via transfer functions. The comparison of selected
inflationary parameters against the observational data, in the case
where the 3-form fields potential have the form $\chi_{i}^{2}+b_{i}\chi_{i}^{4}$,
show that type II solutions, predicting a small variation in the speed
of sound, are in excellent agreement with the observational bounds
of running spectral indexes.

Subsequent tentative lines of exploration, following the herein paper
can be as follows. It would be interesting to explore the generation
of isocurvature perturbations and how they impact on the primordial
non-Gaussianity and the bispectrum. The gravity mediated coupling
through the $\dot{H}$-term in the equations of motion can be explored
in the context of reheating phase after inflation or at the late time
dark energy scenarios with multiple 3-forms \cite{Koivisto:2012xm,Ngampitipan:2011se}.

\section*{Acknowledgments}

Sravan Kumar is grateful for the support of grant SFRH/BD/51980/2012
from Portuguese \foreignlanguage{british}{Agency Funda\c{c}\~ao para
a Ci\^encia e Tecnologia}. This research work was supported by the
grants CERN/FP/123615/2011, CERN/FP/123618/2011, PEst-OE/FIS/UI2751/2014
and PEst-OE/MAT/UI0212/2014. This work was also based on observations
obtained with Planck (http://www.esa.int/Planck), an ESA science mission
with instruments and contributions directly funded by ESA Member States,
NASA, and Canada; and BICEP2 2014 (http://bicepkeck.org./). We also
thanks the COSMOMC team for the resources and software available at
http://cosmologist.info/cosmomc.

\appendix

\section{Appendix: Stability of type I fixed points}

\label{appendix}

Let us now discuss the stability of these fixed points for specific
choice of potentials. The eigenvalues of ${\cal M}_{ij}$ corresponding
to the fixed point $\left(x_{1c},\, w_{1c}\right)$ are $\zeta_{1}=-3$,
$\zeta_{2}=0$. Since the second eigenvalue is zero, we cannot decide
on the stability of this fixed point. The eigenvector for the null
eigenvalue is given by 
\begin{equation}
v_{0}=\left(\begin{array}{c}
\sqrt{2/3}\\
1
\end{array}\right).\label{eigenv-1}
\end{equation}
Let us consider the nonlinear order perturbation in the expansion
\begin{equation}
\delta r'=\mu^{\left(n\right)}\delta r^{n}\label{pertur-eq.}
\end{equation}
where $\delta r=\sqrt{2/3}\delta x_{1}+\delta w_{1}$ is the perturbation
along the direction of the eigenvector (\ref{eigenv-1}). The general
solution of Eq. (\ref{pertur-eq.}) at order $n$ is 
\begin{equation}
\dfrac{\delta r^{\left(-n+1\right)}}{\left(-n+1\right)}=\mu^{\left(n\right)}N+\dfrac{\delta r_{0}^{\left(-n+1\right)}}{\left(-n+1\right)}\quad\textrm{with}\quad\delta r_{0}=\delta r\,\left(N=0\right).\label{pertur-eq-sol}
\end{equation}
For $n>1$, an initial negative perturbation $\delta r_{0}<0$ will
decay if $\mu^{\left(n\right)}$ is positive, with $n$ even, or $\mu^{\left(n\right)}$
is negative and $n$ odd. If the initial perturbation is positive,
then it will decay for $\mu^{\left(n\right)}$ is negative, for all
$n>1$. If we require that $\mu^{\left(1\right)}=1$ in Eq. (\ref{pertur-eq.}),
we must have $\delta x_{1}=\sqrt{3/2}\delta r/2$ and $\delta w_{1}=\delta r/2$.
The procedure consists in evaluating 
\begin{equation}
\delta r'=\sqrt{2/3}\delta x_{1}'+\delta w_{1}'\label{pertur-eq-exp}
\end{equation}
and collecting the second order terms of the expansion of eqs.~(\ref{Dyn-x1})
and (\ref{Dyn-w1}) when the dynamical system is perturbed around
the fixed point 
\begin{eqnarray*}
x_{1} & = & \sqrt{\frac{2}{3}}\cos\theta+\delta x_{1}\\
w_{1} & = & \cos\theta+\delta w_{1}.
\end{eqnarray*}
As the constraint~(\ref{eps -constr-1}) imposes that the dynamical
system, near the fixed point, can only be subjected to a small negative
perturbation, thus, we will consider an initial negative perturbation
$\delta r_{0}<0$. Otherwise, a positive perturbation, that would
slightly increase the value of the two fields above the fixed point,
would imply that the Friedmann constraint (\ref{2Quad-Fried}) would
blow up to infinity.

As it is seen from of eqs.~(\ref{Dyn-x1}) and (\ref{Dyn-w1}), the
presence of the functions $\lambda_{1}$ and $\lambda_{2}$, which,
in turn, depend on the potentials and their derivatives, does not
allow to study in general the stability of the type I solutions. Therefore,
we illustrate this study for some simple and suitable choice of potentials.

\subsection{Identical quadratic potentials}

\label{Quadrid}Let us consider the simple case when the two fields
are under the influence of identical quadratic potentials, i.e., $V\left(x_{1}\right)=x_{1}^{2}$
and $V\left(x_{2}\right)=x_{2}^{2}$. In this situation, Eqs.~(\ref{Dyn-w1})
and (\ref{Dyn-w2}) exhibit type I solutions for any $0<\theta<\pi/2$.
The fixed points for these solutions are constrained by Eq. (\ref{Type I omega}).
Collecting the second order term in (\ref{pertur-eq-exp}) we have
\begin{equation}
\mu^{\left(2\right)}=-\frac{9}{4}\biggl(3\cos\theta+\cos3\theta\biggr),\label{c-quad-quad1}
\end{equation}
which is always negative for $0<\theta\leq\pi/4$. This means that
all fixed points with $0<\theta<\pi/4$ are unstable. If $\theta$
gets larger than $\pi/4$ then the fixed point coordinates $x_{2c}>x_{1c}$
and we can also collect the second order terms in $\delta r'=\sqrt{2/3}\delta x_{2}'+\delta w_{2}'$
for a negative perturbation $x_{2}=\sqrt{\frac{2}{3}}\sin\theta+\delta x_{2}$.
The coefficient yields 
\begin{equation}
\mu^{\left(2\right)}=-\frac{9}{4}\biggl(3\sin\theta-\sin3\theta\biggr),\label{c-quad-quad2}
\end{equation}
which is always negative for $\pi/4\leq\theta<\pi/2$. When the angle
$\theta$ is close to $\pi/2$, then the 3-form field $x_{1}$ approaches
zero and Eq. (\ref{c-quad-quad1}) produces positive values for $\mu^{\left(2\right)}$.
This means that in the asymmetric situation where $x_{1}\approx0$
and $x_{2}\approx\sqrt{2/3}$, the solution $x_{1}\left(N\right)$
converges to zero, however $x_{2}\left(N\right)$ will be unstable.
In fact, from Eq. (\ref{c-quad-quad2}), the second field will eventually
diverge from $\sqrt{2/3}$, when subjected to a small negative perturbation.
Furthermore, the decrease in the value of $x_{2}$ implies that the
variable $w_{2}$ will start to fall faster, as Eq. (\ref{Dyn-x2})
suggests. The decrease of $x_{2}$ will proceed until it reaches zero.
At this point, we can show that both fields will start to oscillate
around zero with a damping factor. The discussion for the situation
where $\theta$ is near zero, is the same, in the sense that the roles
of $x_{1}$ and $x_{2}$, in the previous discussion, are interchanged.
In Fig.~\ref{fig1} (left panel), the behavior of the two fields,
at the end of the inflationary period, when the angle $\theta$ is
close to $\pi/2$, are shown. Therein, we see that the two fields
are going to a damped oscillatory regime, after the divergence of
$x_{2}$ from its fixed point. The herein analytical description is
numerically confirmed.

\subsection{Quadratic and quartic potentials}

\label{Quadr-quartic} When the two fields are subjected to the potentials
$V\left(x_{1}\right)=x_{1}^{2}$ and $V\left(x_{2}\right)=x_{2}^{4}$,
the evolution is generally of the type II. However, Eqs.~(\ref{Dyn-w1})
and (\ref{Dyn-w2}) exhibit type I solutions when the condition (\ref{cond-V1-V2})
holds, which in this case becomes 
\begin{equation}
\left(\frac{1}{\frac{3}{4}\left(\cot\theta\right){}^{2}\csc\theta+\sin\theta}\right)^{2}+\left(\frac{6\cos\theta}{6-\cos2\theta+\cos4\theta}\right)^{2}=1.
\end{equation}
This last condition is satisfied for $\theta\rightarrow\pi/3$ , $\theta\rightarrow\pi/2$
and at $\theta\rightarrow0$. Collecting the second order term in
(\ref{pertur-eq-exp}) we have 
\begin{equation}
\mu^{\left(2\right)}=-\frac{3}{5},
\end{equation}
which is negative. This means that the fixed point with $\theta=\pi/3$
is unstable. At $\theta=0$,i.e, the scenario with the quadratic term
dominance, we must go to third order since, $\mu^{\left(2\right)}=0$.
In that case, collecting the third order terms we have $\mu^{\left(3\right)}=0.28$,
which means that the fixed point is unstable. At $\theta=\pi/2$,
scenario with the quartic term dominance, $\mu^{\left(2\right)}=-7.5$,
which means that the fixed point is unstable.

\bibliographystyle{unsrt}
\bibliography{References}

\end{document}